\documentclass[prl,twocolumn,amsmath,amssymb,floatfix,showpacs,superscriptaddress]{revtex4-2}
\usepackage{graphicx}
\usepackage{amsmath}
\usepackage{placeins}
\usepackage{bm}
\usepackage[dvipsnames]{xcolor}
\usepackage[caption=false]{subfig}
\usepackage{subcaption}
\usepackage{placeins}
\usepackage{ragged2e}
\usepackage{svg}
\usepackage{float}
\usepackage{amsmath}
\DeclareMathOperator{\sgn}{sgn}
\begin{document}

\title{Synchro-nematic and -antinematic ordering of spheroidal circle swimmers}
\author{Anson G. Thambi}
\affiliation{Department of Mechanical Engineering, University of Hawai’i at 
M{\=a}noa, 2540 Dole Street, Holmes Hall 302, Honolulu, Hawaii 96822, USA}

\author{Alexander N. Dodge}
\affiliation{Department of Mechanical Engineering, University of Hawai’i at 
M{\=a}noa, 2540 Dole Street, Holmes Hall 302, Honolulu, Hawaii 96822, USA}

\author{William E. Uspal}
\email{uspal@hawaii.edu}
\affiliation{Department of Mechanical Engineering, University of Hawai’i at 
M{\=a}noa, 2540 Dole Street, Holmes Hall 302, Honolulu, Hawaii 96822, USA}
\affiliation{International Institute for Sustainability with Knotted Chiral Meta Matter (WPI-SKCM\textsuperscript{2}),
1-3-1 Kagamiyama, Higashi-Hiroshima, Hiroshima 739-8526, Japan}

\begin{abstract}
Chirality gives a microswimmer something a straight-line swimmer lacks: a phase. This variable both modulates, and is affected by, the hydrodynamic interactions between microswimmers.   Here we ask what collective order emerges when many such chiral swimmers are free to move, and how the shape and actuation anisotropies of an individual swimmer dictate the outcome. Using a kinetic theory for hydrodynamically interacting circle swimmers, we show that the interplay between intrinsic rotation, stresslet flows, and Jeffery-like reorientation generates effective phase-locking interactions. Asymmetries in the actuation are encoded through a non-axisymmetric stresslet tensor. At the pair level, pusher swimmers select one of two synchronized states depending on particle shape and  actuation asymmetry: in-phase/anti-phase locking, or quarter-shifted  locking. Extending the analysis to many-body systems, we find that these pair-level synchronization mechanisms drive emergent collective phases. The swimmers develop global \textit{synchro-nematic} order when the hydrodynamic coupling favors parallel or anti-parallel phase locking, and \textit{synchro-antinematic} local order where quarter-shifted locking prevails. A coarse-grained field theory predicts the onset of nematic order through a hydrodynamic instability criterion. In addition, we find that the collective states exhibit crystalline or disordered hyperuniform structure arising from period-averaged hydrodynamic interactions that are effectively repulsive between swimmers. Lattice Boltzmann simulations of chiral oblate squirmers, resolving finite-size and near-field flows, recover the synchro-nematic ordering.  Together, these results show how a swimmer's geometric and actuation anisotropies can be leveraged to program synchronization and spatiotemporal order in chiral active matter.
\end{abstract}

\maketitle

\section{Introduction}

Active colloids are systems of micro- or nano-sized particles capable of self-propulsion in a liquid medium through continuous consumption of free energy. This consumption breaks detailed balance at the microscopic level, allowing for collective phases that are inaccessible in thermal equilibrium, including flocks \cite{bricard2013emergence,das2024flocking}, vortices \cite{riedel2005self}, and ``living crystals'' \cite{palacci2013living}. If active colloids have some intrinsic chirality, they can continuously rotate around a body-fixed axis. This introduces an additional source of microscopic symmetry breaking beyond self-propulsion --  broken  mirror (parity) symmetry \cite{lowen2016chirality,liebchen2022chiral,mecke2024emergent}.  This raises a natural question: how does the additional broken symmetry affect collective motion? The consequences are already apparent at the single-particle level: a single spinning colloid will trace a helical or circular trajectory \cite{kummel2013circular}. Thus, chirality introduces a new length scale that typically influences the size and structure of emergent collective states \cite{kummel2013circular,zhang2022hyperuniform,ganguly2023going}. 

This is already evident in ``dry'' models of chiral active particles, where the solvent is not explicitly resolved. The introduction of the orbital radius as a characteristic length scale can allow macroscopic phase separation to be replaced by finite dynamical clusters whose size is set by the underlying orbit \cite{liao2018clustering,ma2022dynamical}. A similar reorganization occurs in polar-aligning systems. Whereas achiral Vicsek-like models form traveling bands and polar flocks, chirality continuously rotates the direction of collective motion, transforming translational order into rotational order. Depending on the strength of chirality, this can produce rotating macro-droplets, micro-flocks, vortices, and cloud-like states, all characterized by coherent circulation rather than straight-line transport \cite{liebchen2017collective}.  In this sense, chirality does not simply modify familiar active matter phases; it reshapes them into structures with intrinsic rotation and often a selected finite size.

Chirality also alters the large-scale fluctuations and transport properties of active matter. Intrinsic rotation can weaken the long-wavelength persistence that underlies Toner--Tu order and giant number fluctuations, opening routes to states with strongly suppressed density fluctuations, including hyperuniform phases \cite{lei2019nonequilibrium}. Beyond pattern formation, chirality can generate odd transport coefficients such as odd viscosity. These coefficients produce transverse responses absent in conventional isotropic fluids, revealing that handed active motion influences not only the structures that emerge, but also the fundamental way in which the material responds to forces and deformations \cite{banerjee2017odd,fruchart2023odd,mecke2023simultaneous}.

The broader motivation is therefore quite plain. Chirality introduces collective states that do not appear in the usual achiral setting \cite{liebchen2017collective,ventejou2021susceptibility,liebchen2022chiral,banerjee2017odd}. It is also not an artificial curiosity. Nature uses rotation across an extraordinary range of scales, from molecular motors such as ATP synthase \cite{noji2001rotary} to circularly swimming algae \cite{huang2021circular}, co-rotating bacteria \cite{petroff2015fast}, spinning embryos \cite{tan2022odd}, and orbiting Volvox colonies \cite{drescher2009dancing}. Synthetic systems offer equally clean routes to chirality, including asymmetric colloids \cite{kummel2013circular}, pear-shaped Quincke rollers \cite{zhang2022hyperuniform}, magnetic spinners \cite{massana2021arrested,shen2023collective,katuri2024control}, and optically driven rotors \cite{kotar2010hydrodynamic}. Chiral active matter is therefore both natural and experimentally accessible.

The story becomes even richer in wet chiral systems, where the surrounding fluid mediates long-ranged interactions between swimmers. For circle swimmers, weak hydrodynamic interactions can accumulate over many revolutions. When circular motion is fast compared with the  relative motion between swimmers, the flow generated over one orbit can be averaged in time, producing an effective interaction that depends on the swimmer type and geometry \cite{michelin2010long,fily2012cooperative}. For example, a pusher-type swimmer, after averaging, behaves like an effective puller along the rotation axis, while in the plane of the orbit its averaged flow is radially outward. Thus, two coplanar pusher circle swimmers experience an effectively repulsive interaction \cite{michelin2010long}. 

This mechanism is evident in the marine alga \textit{E. voratum}, which swim in circles at an air--liquid interface. The resulting effective long-ranged repulsion suppresses large-scale density fluctuations and produces a disordered hyperuniform state \cite{huang2021circular}. In pear-shaped Quincke rollers, vortices and rotating flocks can become hyperuniform \cite{zhang2022hyperuniform}. Driven membrane-bound rotors offer yet another route, where long-ranged hydrodynamic interactions suppress density fluctuations and can even drive the system toward crystalline order \cite{oppenheimer2022hyperuniformity}. Chirality, in short, offers several ways to tame large-scale density noise.

Chirality also brings in the possibility of synchronization. Once a particle rotates, circles, beats, or swims along a helix, there is a phase variable in the problem. The question then becomes whether interactions can lock these phases. In wet systems, hydrodynamic interactions provide a natural route. This is seen in microrotor models, where beads driven around circular tracks synchronize through hydrodynamic coupling \cite{theers2014effects}. The essential ingredient is not merely that the particles move, but that their motion is periodic and that the timing of this periodic motion can be influenced by the fluid flow generated by neighbors. This idea is not restricted to chiral motion.  Hydrodynamic synchronization has been demonstrated in minimal bead-rod swimmer models \cite{putz2009hydrodynamic} and achiral colloidal oscillators \cite{kotar2010hydrodynamic}. Similarly, ``rocking'' Quincke colloids, which move back and forth along a linear axis, develop hydrodynamically mediated phase synchronization together with orientational alignment. This combined phase-and-axis order has been termed \textit{synchronematic order} \cite{leyva2026self}. These examples make an important point: the central ingredient for synchronization is periodicity. Chirality is especially rich because it builds this periodicity directly into the swimmer’s trajectory.

Natural chiral swimmers provide additional realizations of the same broad principle. Sea urchin sperm swimming near surfaces form vortex arrays in which the flagellar beat phase is locked to the angular position around the vortex, generating quantized rotating waves \cite{riedel2005self}. Similarly, spinning \textit{Volvox} colonies form stable hydrodynamic bound states near surfaces, including ``waltz'' and ``minuet'' motions \cite{drescher2009dancing}.  Another  example of hydrodynamic synchronization is found in  pairs of tethered \textit{Bacillus subtilis}. Each tethered bacterium is a living rotor, with a well-defined phase given by its body orientation angle. Through hydrodynamic interactions, a pair of these bacteria can  adjust their angular velocities, lock their phases, and rotate with a common frequency  \cite{oliver2018synchronization}. 

Together, these systems illustrate that hydrodynamic interactions in chiral active matter can produce a range of coordinated states extending beyond phase locking. However, the intricate  details in these systems may tend to obscure generic hydrodynamic mechanisms for spatiotemporal ordering. Surprisingly, however, simple theoretical models for freely moving chiral swimmers that reproduce such ordering are still  nascent in the literature. Most studies of chiral squirmers have focused on single-particle  behavior or pairwise interactions \cite{pedley2016squirmers, burada2022hydrodynamics, maity2022near}. Recently, it was shown that neutral (stresslet-free) spherical squirmers with circular and helical trajectories can spontaneously synchronize their rotation in bulk suspensions. Their swimming directions align and their rotational phases lock, producing both polar order and azimuthal phase order   \cite{samatas2023hydrodynamic}. This naturally raises the question of the effect of the hydrodynamic stresslet, which is expected to generically occur (without fine-tuning) in most active colloidal systems, as well as particle shape. Similarly, the earlier kinetic theory of Ref. \citenum{michelin2010long} left the effect of shape as an open direction of research.

The discussion above sets the stage for the present work. We develop a kinetic theory for freely moving chiral microswimmers based on three generic physical ingredients: intrinsic rotation, stresslet-level hydrodynamic interactions, and geometric shape.  Chirality gives an active particle a built-in rotational phase, while hydrodynamics provides a long-ranged route by which one swimmer can influence the phase and orientation of another. Our approach begins at the pair level. We focus on pusher-type swimmers in the anticipation that effectively repulsive period-averaged flows will maintain large interparticle separations, allowing application of far-field theory. We vary the swimmer geometry from oblate to prolate, and show how non-spherical shape can induce nematic or anti-nematic (quarter-shifted) phase-locking through flow-driven alignment. Furthermore, we find that breaking the axial symmetry of a swimmer's propulsion mechanism can change the character of phase-locking for a given swimmer geometry. We then broaden our view to monolayers of many swimmers, where microscopic synchronization can feed into macroscopic order. We find that pairwise synchronization can lead to collective states with orientational, spatial, and temporal ordering. 

Overall, we integrate chirality, hydrodynamics, and particle shape in a single theoretical framework. Our approach permits resolution of how collective phases emerge from the circular motion and interaction of individual, freely moving chiral swimmers, and how this emergence is shaped by the  microhydrodynamic character of the swimmer itself. As a result, our study sheds light on how the basic microscopic design of an individual circle swimmer (shape, actuation) can be leveraged to induce spatially and temporally structured collective phases.

\section{Model}

\subsection{Kinetic theory}

To study the collective motion of many active particles, we require a model that keeps the essential hydrodynamic physics while remaining simple enough to resolve large systems over long times. We therefore adapt the equations of motion developed by Saintillan and Shelley for a self-propelled particle in a locally linear or linearized flow \cite{saintillan2008instabilities, saintillan2018rheology, thambi2025clustering}. In this description, each swimmer generates an active stresslet, and the motion of every other swimmer is influenced by the velocity field associated with these stresslets. The active stresslet encodes the leading-order term in a far-field expansion for the fluid flow sourced by the swimming activity of a particle. Therefore, our model truncates activity-sourced flow at leading order in interparticle separation. 

In many experiments on active colloids, the particles sediment close to the bottom substrate and move predominantly in a plane. We therefore restrict the swimmer centers to the $xy$ plane. The position of swimmer $i$ is denoted by $\mathbf{x}_i$, and its swimming direction by the unit vector $\hat{\mathbf{d}}^{\,(i)}$. The quantity $\hat{\mathbf{d}}^{\,(i)}$ is also assumed to be restricted to the $xy$ plane. This restriction is naturally realized by rod-like particles that have sedimented to a bottom surface. For spheres, or for discoidal particles that have $\hat{\mathbf{d}}^{\,(i)}$ oriented along the minor axis, this restriction is straightforwardly realized in systems of metallo-dielectric Janus particles that move by induced charge electrophoresis \cite{katuri2022arrested}.  Additionally, catalytic Janus spheres typically approximately align $\hat{\mathbf{d}}^{\,(i)}$ in the plane of the substrate through phoretic and hydrodynamic interactions with the substrate \cite{uspal2015self}.

The translational velocity of the swimmer is written as
\begin{equation}
    \mathbf{U}^{(i)}
    =
    U_s \hat{\mathbf{d}}^{\,(i)}
    +
    \mathbf{u}(\mathbf{x}_i),
    \label{eq:particle_vel}
\end{equation}
where $U_s$ is the self-propulsion speed. The first term is the swimmer's own motion, while $\mathbf{u}(\mathbf{x}_i)$ is the hydrodynamic velocity at its position $\mathbf{x}_i$ due to all the other swimmers.

The orientational dynamics are determined by how the swimmer responds to the local ambient flow. The local velocity gradient may rotate the swimmer through vorticity and may also reorient it through the rate of strain. In addition, the swimmers considered here are circular swimmers: they possess an intrinsic angular velocity, denoted by $\Omega_s$. This rotation is not imposed by an external torque, as in magnetically driven chiral colloids; rather, it arises from the particle's own geometry or actuation pattern. Consequently, no rotlet is included in the present far-field description. The angular velocity of swimmer $i$ is then written as a Jeffery equation \cite{lauga2020fluid}
\begin{equation}
    \dot{\theta}^{(i)}
    =
    \Omega_s
    +
    \hat{\mathbf{c}}^{\,(i)}
    \cdot
    \left[
        \Gamma_i \mathbf{E}(\mathbf{x}_i)
        +
        \mathbf{W}(\mathbf{x}_i)
    \right]
    \cdot
    \hat{\mathbf{d}}^{\,(i)}.
    \label{eq:jeffery}
\end{equation}
Here, $\hat{\mathbf{c}}^{\,(i)}$ is the in-plane unit vector perpendicular to $\hat{\mathbf{d}}^{\,(i)}$. The tensors $\mathbf{E}$ and $\mathbf{W}$ are the symmetric and antisymmetric parts of the velocity gradient,
\begin{equation}
    \mathbf{E}(\mathbf{x})
    =
    \frac{1}{2}
    \left[
        \nabla \mathbf{u}
        +
        \left(\nabla \mathbf{u}\right)^{T}
    \right],
    \label{eq:strain}
\end{equation}
and
\begin{figure*}[hbtp]
    \centering
    \includegraphics[width=\textwidth]{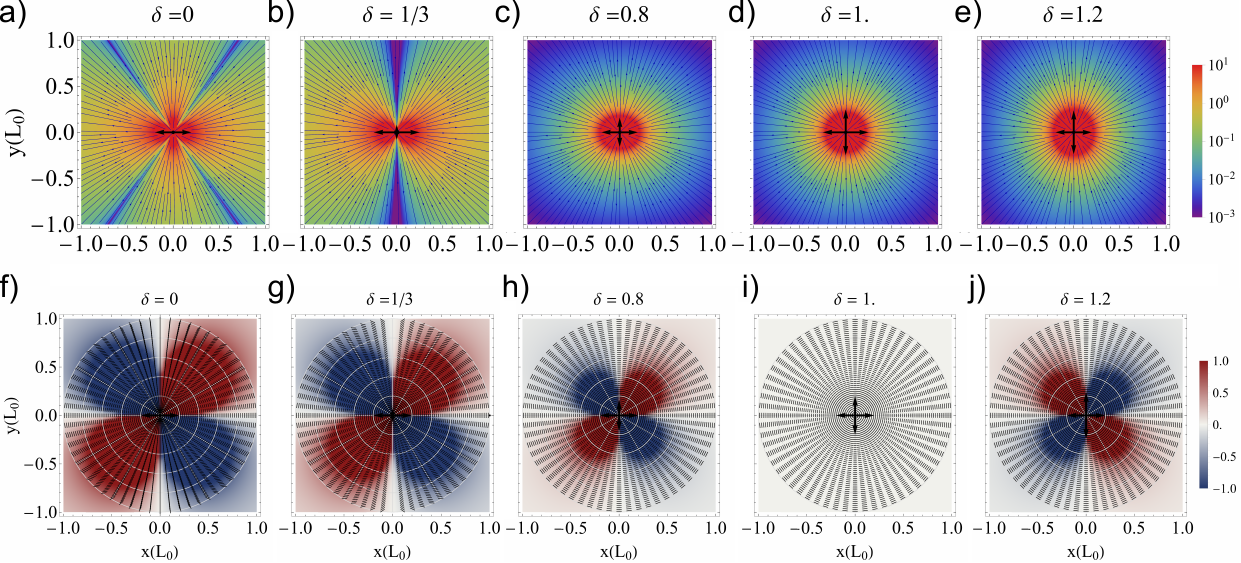}
    \caption{\label{fig:flowfield}\justifying{Flow, strain and vorticity fields generated by a non-axisymmetric pusher stresslet placed at the origin and oriented along $\hat{\mathbf{x}}$. (a)--(e) Dimensionless velocity magnitude, $|\mathbf{u}|/U_0$, for $\delta=0,1/3,0.8,1, 1.2$, with stresslet strength $\sigma_0/(\mu L_0^2 U_0)=-6$. Double-headed arrows represent the stresslet terms in Eq. \ref{eq:three_stresslets}. (f)--(j) Corresponding vorticity fields, in units of $U_0/L_0$, with black line segments marking the local elongation axis of the rate-of-strain tensor. Across all values of $\delta$, the elongation axes retain nematic symmetry, favoring the alignment of the major axis of a neighboring swimmer either parallel or antiparallel to the local stretching direction. The vorticity changes sign as $\delta$ crosses unity and vanishes at $\delta=1$, marking the transition between the two rotational flows.}}
\end{figure*}
\begin{equation}
    \mathbf{W}(\mathbf{x})
    =
    \frac{1}{2}
    \left[
        \nabla \mathbf{u}
        -
        \left(\nabla \mathbf{u}\right)^{T}
    \right].
    \label{eq:vorticity}
\end{equation}
The parameter $\Gamma_i$ is the Jeffery shape parameter,
\begin{equation}
    \Gamma_i
    =
    \frac{1-r_e^2}{1+r_e^2},
    \label{eq:gamma}
\end{equation}
where $r_e^{-1}$ is the ratio of the semi-axis of symmetry of the spheroid to either of the two remaining semi-axes. Thus, $\Gamma=0$ corresponds to a sphere, $\Gamma>0$ to a prolate spheroid, and $\Gamma<0$ to an oblate spheroid.

We now specify the hydrodynamic flow field. At distances large compared with the particle size, the leading contribution to the flow generated by a force- and torque-free swimmer is the active stresslet. This far-field approximation neglects the near-field details of the flow, but retains the dominant long-ranged hydrodynamic disturbance through which the swimmers interact. The velocity induced at the position of swimmer $i$ by all other swimmers is written as
\begin{equation}
    \mathbf{u}(\mathbf{x}_i)
    =
    \sum_{j\neq i}^{N}
    \frac{3}{8\pi\mu}
    \frac{ (\mathbf{x}_{j} - \mathbf{x}_{i})
        \big((\mathbf{x}_{j} - \mathbf{x}_{i})
            \cdot
            \mathbf{S}_{j}
            \cdot
            (\mathbf{x}_{j} - \mathbf{x}_{i})
        \big)
    }{
        |\mathbf{x}_{j} - \mathbf{x}_{i}|^{5}
    }.
    \label{eq:stresslet_flow}
\end{equation}
Here, $\mu$ is the dynamic viscosity of the fluid, $N$ is the number of swimmers, and $\mathbf{S}_j$ is the stresslet tensor of swimmer $j$. 

For an axisymmetric squirmer, the stresslet has the coordinate-free form \cite{saintillan2018rheology}
\begin{equation}
    \mathbf{S}_0
    =
    \sigma_0
    \left(
        \hat{\mathbf{d}}\hat{\mathbf{d}}
        -
        \frac{\boldsymbol{\mathcal{I}}}{3}
    \right).
    \label{eq:axis_stresslet}
\end{equation}
The sign of $\sigma_0$ sets the hydrodynamic character of the swimmer: $\sigma_0<0$ corresponds to a pusher, while $\sigma_0>0$ corresponds to a puller.

The above form assumes that the two directions perpendicular to the swimming axis are hydrodynamically equivalent. This is no longer the case when the swimmer has a non-axisymmetric slip distribution \cite{nishiguchi2015mesoscopic,poehnl2023shape}. To include this effect, we add a symmetry-breaking contribution to the stresslet~
\begin{equation}
    \mathbf{S}
    =
    \mathbf{S}_0
    +
    \sigma_0 \delta
    \left(
        \hat{\mathbf{c}}\hat{\mathbf{c}}
        -
        \hat{\mathbf{e}}\hat{\mathbf{e}}
    \right).
    \label{eq:non_axis_stresslet}
\end{equation}
The vectors $\hat{\mathbf{c}}$, $\hat{\mathbf{d}}$, and $\hat{\mathbf{e}}$ define the body-fixed principal axes of the stresslet. We take $\hat{\mathbf{d}}$ to be aligned with the swimming direction, while $\hat{\mathbf{c}}$ and $\hat{\mathbf{e}}$ span the two directions transverse to it. In the quasi-two-dimensional setting used here, $\hat{\mathbf{c}}$ and $\hat{\mathbf{d}}$ are confined to the $xy$ plane, and the remaining axis is fixed by the right-handed convention $\hat{\mathbf{e}}=\hat{\mathbf{c}}\times\hat{\mathbf{d}}$. The intrinsic angular velocity is therefore directed normal to the plane of motion, along $\hat{\mathbf{e}}$, so that its vector form may be written as $\boldsymbol{\Omega}_s=\Omega_s\hat{\mathbf{e}}$, with the sign of $\Omega_s$ setting the sense of rotation.

The dimensionless parameter $\delta$ measures how strongly the stresslet departs from axisymmetry. When $\delta=0$, the swimmer is hydrodynamically indifferent to the two transverse directions. When $\delta\neq0$, the $\hat{\mathbf{c}}\hat{\mathbf{c}}$ and $\hat{\mathbf{e}}\hat{\mathbf{e}}$
contributions are unequal,  which is the expected hydrodynamic signature  from a particle with non-axisymmetric slip. The form in Eq.~\eqref{eq:non_axis_stresslet} remains traceless, $\operatorname{tr}(\mathbf{S})=0 $, which follows from incompressibility of the flow.

Hence, to sum up, each swimmer propels along its body axis, turns with an intrinsic angular velocity $\Omega_s$, and generates a long-ranged stresslet flow. At the same time, it is advected and reoriented by the collective flow produced by all the other swimmers. The coupled translational and rotational dynamics are therefore obtained from Eqs.~\eqref{eq:particle_vel} and \eqref{eq:jeffery}, with the hydrodynamic velocity given by Eq.~\eqref{eq:stresslet_flow}. These coupled equations are integrated numerically using an explicit Euler scheme.

All swimmers are assumed to have identical parameters in a given simulation. The simulations are performed in a periodic square domain with side lengths $L_x=L_y$. The minimum image convention is assumed (each particle interacts only with the nearest periodic image of every other particle) \cite{thambi2025clustering}. The particle semi-major axis defines the characteristic length scale $L_0$, while $U_0$ is used as the characteristic velocity scale. The corresponding time scale is $ T_0={L_0}/{U_0}$. Throughout this paper, we set ${U_s}/{U_0}=0.5$.

\subsection{The squirmer model and the Lattice Boltzmann Method}

The model equations developed above are ``point-particle'' equations from a hydrodynamic point of view because a boundary condition for the flow field is not explicitly imposed on the surface of each particle. Secondly, in our kinetic equations, the velocity field sourced by an active particle is truncated at the leading order far-field term. Higher order, faster decaying hydrodynamic interactions are not included, although these could be important for particles near contact. 

To resolve hydrodynamic finite-size and near-field effects, we also consider the so-called ``squirmer model'' \cite{lighthill1952squirming, blake1971spherical, ishikawa2024fluid}.  In this model, the fluid velocity $\mathbf{u}(\mathbf{x})$ and pressure $p(\mathbf{x})$ satisfy the Stokes equation $-\nabla p + \mu \nabla^2 \mathbf{u} = 0$ and incompressibility condition $\nabla \cdot \mathbf{u} = 0$. A hydrodynamic boundary condition $\mathbf{u}(\mathbf{x}) = \mathbf{U}^{(i)} + \bm{\Omega}^{(i)} \times (\mathbf{x} - \mathbf{x}_i) + \mathbf{v}_{s}^{(i)}(\mathbf{x})$ is imposed on the surface of each particle $i$, where $\mathbf{U}^{(i)}$ and $\bm{\Omega}^{(i)}$ are the translational and angular velocities, respectively, and $\mathbf{x}_i$ is position of the particle centroid. The so-called slip velocity $\mathbf{v}_{s}^{(i)}(\mathbf{x})$ provides the interfacial actuation that powers self-propulsion. The slip velocity is typically expanded in an orthonormal basis of so-called squirming modes, with a prescribed set of squirming mode amplitudes $B_n$ with $n \in \{1, 2, \ldots\}$ (see below).  In the absence of external forces or torques, the particles are individually force-free and torque-free. These conditions close the system of equations for the unknowns $\mathbf{U}^{(i)}$ and $\bm{\Omega}^{(i)}$.

Although the squirmer model was originally developed for spherical particles, extensions to spheroidal particles have been presented in the literature \cite{ishikawa2006interaction,leshansky2007frictionless, felderhof2016stokesian, theers2016modeling, poehnl2020axisymmetric}. In particular,  Ref. \citenum{poehnl2020axisymmetric} developed a complete and orthonormal set of axisymmetric squirming modes for both oblate and prolate spheroidal particles. For an oblate spheroid, the slip is written in modified oblate spheroidal coordinates $(\lambda, \zeta, \phi)$ (see Appendix A) as 
\begin{equation}
v_s(\zeta) = \lambda_0 \sum_{n \geq 1} B_n V_n(\zeta).  
\label{eq:vs_LBM}
\end{equation}
Here, $\lambda = \lambda_0$ defines the surface of the particle, and $V_n(\zeta) = (\lambda_0^2 + \zeta^2)^{-1/2} \, P_n^1(\zeta)$, where $P_n^1(\zeta)$ is an associated Legendre polynomial. One of the main findings of Ref. \citenum{poehnl2020axisymmetric} is that the odd mode amplitudes contribute to the particle velocity, while the even mode amplitudes contribute to the stresslet strength, through linear coefficients that depend on the particle aspect ratio $r_e$. Therefore, truncating the slip to the first two squirming modes is sufficient to realize self-propulsion and a particle-sourced flow field that includes a force-dipole (stresslet) term.

To our knowledge, circling motion of oblate spheroidal squirmers has not previously been studied in the literature. Here, in order to realize (torque-free) circle swimmers, we introduce an oblate chiral squirming mode in the following way. We consider a $B_1$ mode and select a plane containing the minor axis and a major axis. One side of this plane, we negate the slip velocity. The amplitude of this chiral squirming mode is denoted $B^*$.

In order to solve the squirmer model for a system of $N_s$ interacting oblate spheroidal squirmers, we use the D3Q19 Lattice Boltzmann Method (LBM) with a single relaxation time (BGK) operator to model the hydrodynamics of the suspending fluid. Spherical \cite{alarcon2017morphology, kuron2019lattice} and prolate \cite{thampi2024simulating, mishra2026interface} squirmers have been implemented in the LBM in previous research. Our implementation of LBM is standard. We briefly discuss the implementation in Appendix A. Details concerning the LBM are presented extensively elsewhere \cite{ladd2001lattice,aidun2010lattice}.

We note that in our LBM simulations, the particles are confined between parallel solid walls, and swim close to the bottom wall. This confinement is expected to screen long-ranged hydrodynamic interactions between particles \cite{liron1976stokes}. Therefore, close quantitative agreement between the LBM results and the kinetic theory is not expected. On the other hand, qualitative agreement between the two approaches would demonstrate that shape-induced synchronization and ordering is robust against confinement. 

We validate our scheme by considering the velocity $U_s$ of a single squirmer between parallel walls (Appendix A). We obtain good agreement with the numerical and analytical results of Poehnl \textit{et al.} \cite{poehnl2020axisymmetric}, especially for smaller aspect ratios. Appendix A also presents the relationship between the mode amplitude $B^*$ and the angular velocity for different aspect ratios $r_e$.

\section{Results and discussion}

\subsection{Flow fields}
\begin{figure*}[hbtp]
    \centering
    \includegraphics[width=\textwidth]{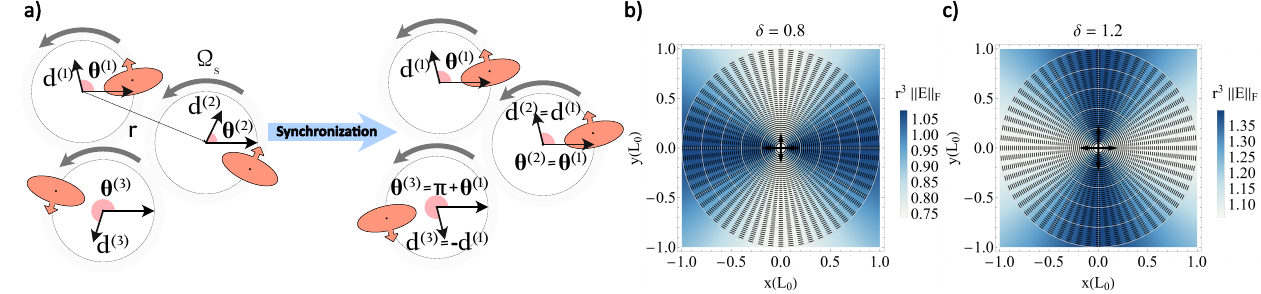}
    \caption{\label{fig:schematic}\justifying{(a) Schematic of hydrodynamic synchronization between freely swimming chiral pusher swimmers. The left side shows the initial state, illustrated here for oblate pushers shown in orange. The swimmers are randomly positioned and oriented, with body axes $\mathbf{d}^{(i)}$, translational speed $U_s$, and intrinsic angular velocity $\Omega_s$, so that each swimmer follows a circular trajectory of radius $R_0=U_s/\Omega_s$. The separation between neighboring orbit centres is taken to satisfy $r\gg R_0$. In this limit, the fast circular motion can be averaged out and each swimmer can be represented by its orientation vector placed at the centre of its orbit. The right side shows the synchronized state: neighboring swimmers lock into nematic configurations, aligning either parallel or antiparallel to one another, such that $\mathbf{d}^{(2)}=\mathbf{d}^{(1)}$ and $\mathbf{d}^{(3)}=-\mathbf{d}^{(1)}$.} (b-c) Scaled magnitude (Frobenius norm) $r^3 || \mathbf{E}(\mathbf{x}) ||_{F}$ of the strain field sourced by a non-axisymmetric stresslet for (b)  $\delta = 0.8$ and (c) $\delta = 1.2$. The quantity $\hat{\mathbf{d}}$ is oriented in the positive x-direction.}
\end{figure*}
To set the stage for consideration of hydrodynamic interactions between swimmers, we first consider the flow sourced by a single swimmer. In Fig. \ref{fig:flowfield}(a-e), we show the streamlines and flow magnitude $|\mathbf{u}(\mathbf{x})|$ in the xy plane for a pusher ($\sigma_0 < 0$) stresslet located at the origin. The propulsion direction $\hat{\mathbf{d}}$ is oriented in the positive x-direction, and therefore the transverse direction $\hat{\mathbf{c}}$ is oriented in the negative y-direction. The streamlines for an axisymmetric pusher ($\delta = 0$) have a highly anisotropic character: flow is pumped away from the swimmer along the propulsion axis, and drawn into the swimmer along the transverse axis. 

For $\delta \neq 0$, it is instructive to re-write  Eq. \eqref{eq:non_axis_stresslet} as
\begin{equation}
\mathbf{S} = \sigma_0
    \left(
        \hat{\mathbf{d}}\hat{\mathbf{d}}
        -
        \frac{\boldsymbol{\mathcal{I}}}{3}
    \right)
     +      
\sigma_0 \delta
    \left(
        \hat{\mathbf{c}}\hat{\mathbf{c}}
        -
        \frac{\boldsymbol{\mathcal{I}}}{3}
    \right)
    -
    \sigma_0 \delta
    \left(
        \hat{\mathbf{e}}\hat{\mathbf{e}}
        -
        \frac{\boldsymbol{\mathcal{I}}}{3}
    \right).
    \label{eq:three_stresslets}
\end{equation}
Therefore, setting $\delta \neq 0$ can be regarded as introducing additional stresslets aligned with the two transverse axes (in-plane and vertical). Schematically, the in-plane transverse stresslet is shown as a second double-headed arrow in Fig. \ref{fig:flowfield}(b-e). As $\delta$ is increased, the region of inflow shrinks, and is finally eliminated at $\delta = 1/3$. With further increase of $\delta$ in the range $1/3 < \delta < 1$, the flow assumes more and more of an isotropic character. At $\delta = 1$, the flow is completely isotropic in the xy plane, as a consequence of the identical strengths of the two in-plane stresslets. For $\delta > 1$, the flow is once again anisotropic, with stronger magnitude in the transverse direction. 

Now we consider the two components of the flow gradient, as they are responsible for orientational interactions. Fig. \ref{fig:flowfield}(f-j) show the vorticity (background color) and the local elongation axes (black line segments). According to its sign, the vorticity rotates a particle counterclockwise (positive) or clockwise (negative). Concerning the extensional axis, a spheroidal particle tends to align its major axis with it. Accordingly, a prolate particle will tend to align its propulsion direction to be parallel or anti-parallel with the black line segments. An oblate particle, on the other hand, will tend to align its propulsion direction to be transverse to the line segments. Importantly, this orientational interaction has nematic (fore-aft) symmetry.

Notably, the vorticity vanishes at $\delta = 1$, and switches sign as $\delta$ crosses this value. The pattern defined by the local axes of extension is highly anisotropic at $\delta = 0$, and has an approximately dipolar character. However, as $\delta$ is increased, these line segments become approximately azimuthal (\textit{i.e.}, tangent to circles centered on the origin).

\subsection{Pairwise interactions and phase-locking}

Now we consider whether the flow produced by one swimmer influences the other in a way that favors orientational alignment. For oblate swimmers, the rate-of-strain fields shown in Fig.~\ref{fig:flowfield}, especially the nearly azimuthal axes of extension (black line segments) in Fig.~\ref{fig:flowfield}h-j, provide a reason to think that it does. The tendency to align the particle's major axis with the local axis of extension  corresponds to a tendency to align the propulsion axis of an oblate swimmer with the center-to-center axis, although without distinguishing between parallel and anti-parallel orientations. This promotes  nematic-like bound configurations, as previously explored for linear swimmers in Ref. \cite{thambi2025clustering}. This is clearly a nonreciprocal orientational interaction, as can be seen from the case in which one swimmer is aligned with the center-to-center axis and the other is not.  

Here, we ask what becomes of this mechanism when the swimmers are intrinsically rotating. If the strain-induced tendency towards alignment or anti-alignment persists during rotation, then nearby swimmers can potentially lock their relative phase. This is important, since then hydrodynamic interaction would provide a route to synchronization. We therefore begin with the simplest scenario: two interacting circle swimmers. From this pair problem, we derive an effective equation for the slow (compared to the orbital period) variation of the phase difference, which allows us to test whether the stresslet flow can generate phase locking, before turning to collective behavior.
\begin{figure*}[hbtp]
    \centering
    \includegraphics[width=\textwidth]{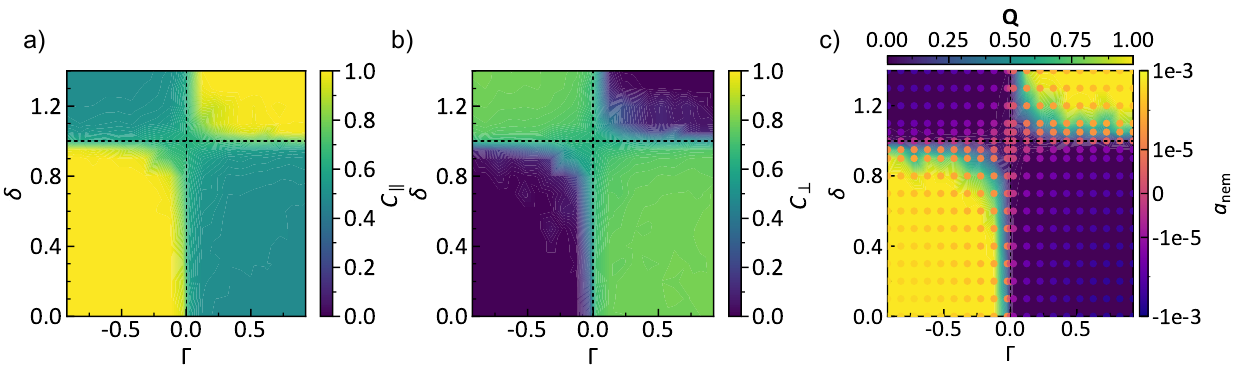}
    \caption{\label{fig:phase_diagram}\justifying{Local phase locking and collective nematic order in the $(\Gamma,\delta)$ plane. (a) Nearest-neighbour parallel-locking measure $C_{\parallel}=\langle |\cos\Delta_{ij}|\rangle_{\mathrm{nn}}$, computed for neighbour pairs within $20 \, L_0$ in simulations with $N=1000$ and $\phi=0.005$. Large values indicate local parallel or antiparallel alignment. (b) Corresponding quarter-phase measure $C_{\perp}=\langle |\sin\Delta_{ij}|\rangle_{\mathrm{nn}}$, with large values indicating local locking near $\Delta_{ij}=\pi/2$ or $3\pi/2$. The black dashed lines mark the pair-level stability boundaries predicted by Eqs.~\eqref{eq:phase_stability_parallel} and \eqref{eq:phase_stability_quarter}. (c) Phase diagram of the global nematic order. The nematically ordered regions agree with the continuum prediction, oblate pushers order for $\delta<1$, while prolate pushers order for $\delta>1$. The overlap between the high-$C_{\parallel}$ regions in panel (a) and the nematic regions in panel (c) shows that the collective nematic phase grows out of the local pair-locking mechanism. The growth rate is calculated from Eq. \eqref{eq:nematic_growth_final}, with the correlation function $g$ determined from the numerical simulations.}}
\end{figure*}

Specifically, we consider two widely separated circle swimmers and define their phase difference as
\begin{equation}
    \Delta=\theta^{(2)}-\theta^{(1)}.
    \label{eq:phase_difference}
\end{equation}
The phase dynamics follows from
\begin{equation}
    \dot{\Delta}=\dot{\theta}^{(2)}-\dot{\theta}^{(1)} .
    \label{eq:phase_difference_rate}
\end{equation}
Using the Jeffery equation for each swimmer, Eq. \eqref{eq:jeffery}, the intrinsic angular velocities cancel in Eq.~\eqref{eq:phase_difference_rate}, since the two swimmers have the same $\Omega_{\mathrm{s}}$. Thus the slow
evolution of $\Delta$ is controlled entirely by the hydrodynamic rotations induced by the stresslet flows. The swimmers move on circular trajectories, but when the distance between their orbit centers is
large compared with the orbital radius, their positions may be approximated as fixed at the centres of their orbits. We therefore place swimmer 1 at the origin and swimmer 2 at
\begin{equation}
    \mathbf{x}_2=(x,y,0),
    \qquad
    r=\sqrt{x^2+y^2}.
\end{equation}
The orientations continue to rotate, and we write
\begin{equation}
    \theta^{(2)}=\theta^{(1)}+\Delta,
    \label{eq:theta2_theta1_delta}
\end{equation}
where $\theta^{(1)}$ is the fast phase and $\Delta$ is treated as slowly varying.

The hydrodynamic angular velocity of swimmer 2 due to swimmer 1 has the form
\begin{equation}
    \dot{\theta}^{\,(2)}_{\mathrm{hydro}}
    =
    \hat{\mathbf{c}}^{\,(2)}\cdot
    \left[
        \Gamma \mathbf{E}_{1\to2}
        +
        \mathbf{W}_{1\to2}
    \right]
    \cdot
    \hat{\mathbf{d}}^{\,(2)}.
    \label{eq:theta2_hydro}
\end{equation}
Here $\mathbf{E}_{1\to2}$ and $\mathbf{W}_{1\to2}$ are the rate-of-strain and vorticity tensors generated by swimmer 1 and evaluated at the position of swimmer 2. Since phase locking is a slow process, we average this expression over one period of the fast phase $\theta^1$ while holding $\Delta$ fixed:
\begin{equation}
    \left\langle
    \dot{\theta}^{\,(2)}_{\mathrm{hydro}}
    \right\rangle
    =
    \frac{1}{2\pi}
    \int_0^{2\pi}
    \dot{\theta}^{\,(2)}_{\mathrm{hydro}}
    \,d\theta^1 .
    \label{eq:phase_average}
\end{equation}

Carrying out this average gives a simple result. The contribution from the orientation-averaged vorticity vanishes, while the rate-of-strain contribution survives:
\begin{equation}
    \left\langle
    \dot{\theta}^{\,(2)}_{\mathrm{hydro}}
    \right\rangle
    =
    -
    \frac{3\Gamma\sigma_0(1-\delta)}
    {64\pi \mu r^3}
    \sin(2\Delta).
    \label{eq:theta2_hydro_avg}
\end{equation}

By symmetry, the correction to swimmer 1 is obtained by replacing $\Delta\to-\Delta$. Substituting into \eqref{eq:phase_difference_rate} gives the effective slow phase equation
\begin{equation}
    \dot{\Delta}
    =
    -
    \frac{3\Gamma\sigma_0(1-\delta)}
    {32\pi \mu r^3}
    \sin(2\Delta).
    \label{eq:effective_phase_equation}
\end{equation}
Thus, the coupling is proportional to $\Gamma$, so the effect vanishes for particles that do not respond to the rate of strain. It is also proportional to the stresslet strength $\sigma_0$, and is modified by the non-axisymmetry through the factor $(1-\delta)$. Finally, the dependence on $\sin(2\Delta)$ shows that the coupling is nematic in phase, \textit{i.e.}, it does not 
distinguish between synchronized and anti-synchronized orientations.
The fixed points where $\dot{\Delta} = 0$ are
\begin{equation}
    \Delta=0,\quad
    \frac{\pi}{2},\quad
    \pi,\quad
    \frac{3\pi}{2}.
\end{equation}
Hence, for a pusher swimmer $(\sigma_0 < 0)$ considered here, linear stability gives
\begin{equation}
    \Delta=0,\pi
    \quad \text{stable if} \quad
    \Gamma(1-\delta)<0.
    \label{eq:phase_stability_parallel}
\end{equation}
This is \textit{in-phase/anti-phase locking} or \textit{nematic locking}. Furthermore, we  obtain 
\begin{equation}
    \Delta=\frac{\pi}{2},\frac{3\pi}{2}
    \quad \text{stable if} \quad
    \Gamma(1-\delta)>0.
    \label{eq:phase_stability_quarter}
\end{equation}
which is \textit{quarter-shifted locking} or \textit{anti-nematic locking}.

Thus, the stresslet-mediated interaction produces synchronization between widely separated circular swimmers. Which phase-locked state is selected is controlled by the sign of $\Gamma(1-\delta)$.

\subsection{Physical mechanism of phase-locking}

At this point, it is natural to ask how these results are related to the flow fields shown in Fig. \ref{fig:flowfield}. Consider prolate pushers with $\delta = 1.2$. From Eq. \eqref{eq:phase_stability_parallel}, we expect a tendency towards parallel/anti-parallel alignment. Now we turn to Fig. \ref{fig:flowfield}j to try to develop physical intuition for this tendency.  If swimmer 2 is located on the y-axis, the local rate-of-strain does indeed promote parallel/anti-parallel alignment. However, if swimmer 2 is located on the x-axis, the local rate-of-strain promotes a transverse orientation. How can the first tendency predominate? Here, it must be borne in mind that both swimmers are rotating. If swimmer 2 is located on the x-axis of Fig. \ref{fig:flowfield}j, it will initially experience a tendency towards transverse alignment, but at a later time, \textit{e.g.}, after a $90^{\circ}$ rotation of both swimmers, it will experience a tendency towards parallel/anti-parallel alignment. (To simplify the argument, we assume the spatial location of swimmer 2 relative to swimmer 1 is  fixed.) We also observe that if swimmer 2 is initially oriented parallel to swimmer 1, its major axis is initially under compression, and is later under extension.

Given that swimmer 2 experiences both orientational tendencies, the question of which one prevails depends on the spatial variation of the magnitude of the rate-of-strain. Fig. \ref{fig:schematic}(b-c) shows the Frobenius norm of the rate-of-strain for $\delta = 0.8$ and $\delta = 1.2$. This quantity is \textit{not} radially symmetric. It is clear that, for prolate particles, the transverse aligning tendency prevails for $\delta = 0.8$, and the parallel/anti-parallel aligning tendency prevails for $\delta = 1.2$. Straightforwardly, the opposite holds for oblate swimmers. 

As further mathematical support of these arguments, in Appendix B, we present a detailed analysis of the vorticity and rate-of-strain in the rotating body frame of swimmer 2, averaged over one period of rotation. We show that (i) the period-averaged effect of vorticity vanishes, as mentioned previously, and (ii) the effective, period-averaged axes of compression and extension in the body frame are swapped as $\delta$ crosses $\delta = 1$.

\begin{figure*}[hbtp]
    \centering
    \includegraphics[width=\textwidth]{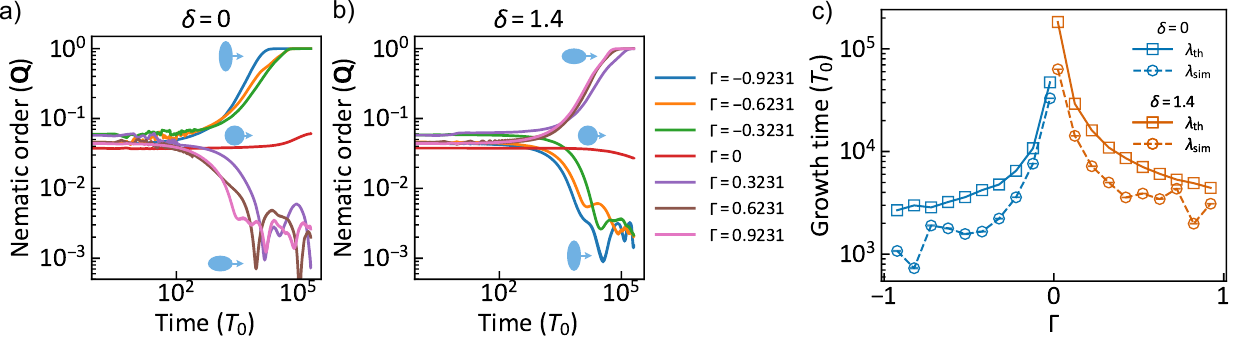}
    \caption{\label{fig:nematic}\justifying{Time evolution and growth rate of nematic ordering for chiral pusher swimmers. (a) Global nematic order $Q$ as a function of time for pusher-type axisymmetric stresslets ($\delta=0$) across different values of $\Gamma$, spanning oblate to prolate swimmer shapes. For axisymmetric actuation, nematic order develops only for oblate swimmers. (b) Corresponding evolution for a non-axisymmetric stresslet with $\delta=1.4$. In this case, the ordering regime shifts, and only prolate pushers develop nematic order. (c) Early-time exponential fits of the nematic growth for $\delta=0$ and $\delta=1.4$, used to extract the simulated growth rate $\lambda_{\mathrm{sim}}$. The fitted values are compared with the theoretical prediction $\lambda_{\mathrm{th}}$ from Eq.~\eqref{eq:nematic_growth_final}, with the correlation function $g$ determined from the numerical simulations. The agreement shows that stronger flow-alignment response, controlled by $\Gamma$, accelerates the onset of nematic order in the regimes selected by the stability condition.}}
\end{figure*}

\subsection{Phase diagram}
Having established the phase locking at the two-swimmer level, we next ask whether the same mechanism leaves its mark on the many-body dynamics. We therefore simulate monolayers at fixed number density
$\phi={N L_0^2}/{L_x^2}=0.005$ while varying the particle shape parameter $\Gamma$ from oblate to prolate and
the non-axisymmetry parameter $\delta$. Throughout these simulations we set $U_s/\Omega_{\mathrm{s}}=L_0$, such that the orbital radius $R_0 = L_0$. 

\begin{figure*}[hbtp]
    \centering
    \includegraphics[width=\textwidth]{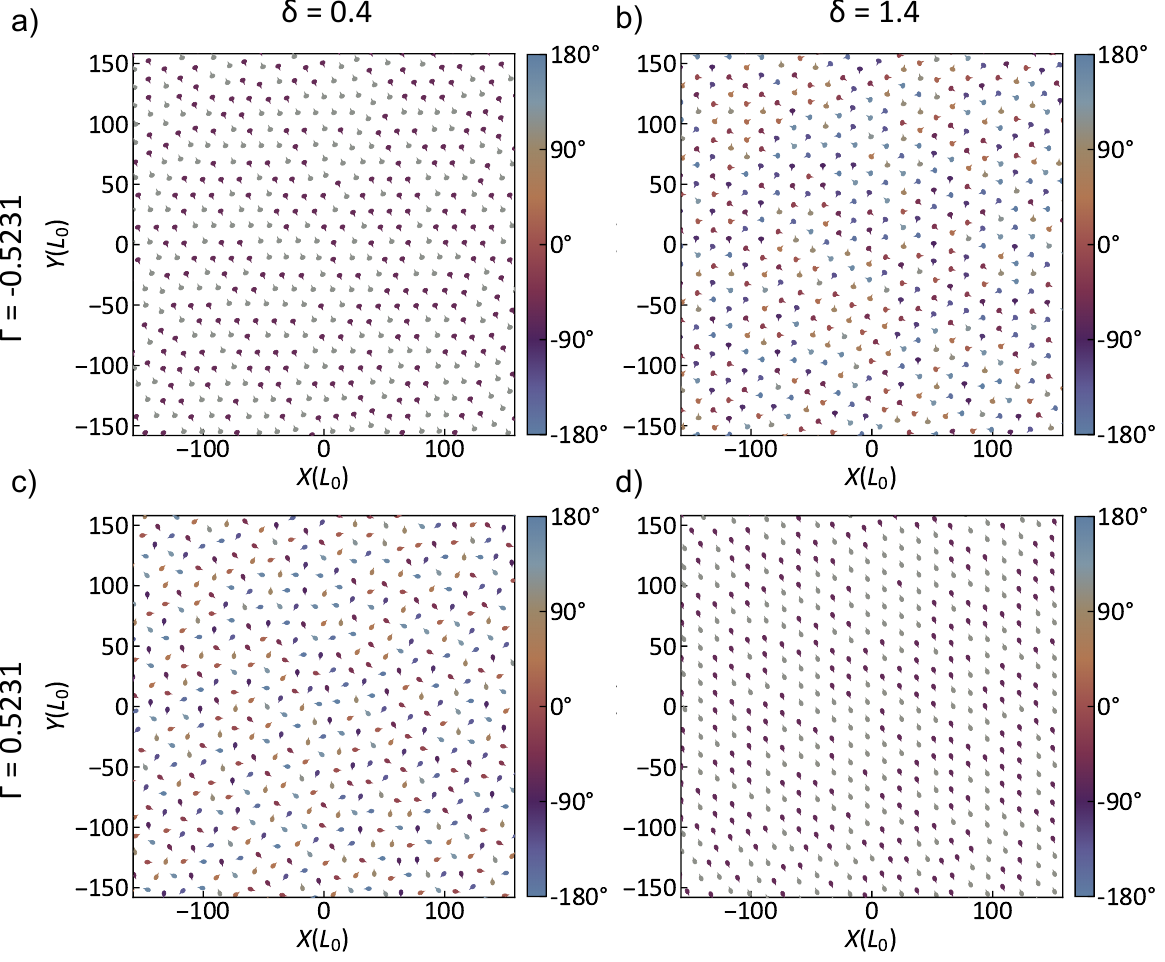}
    \caption{\label{fig:snapshots}\justifying{Steady-state particle configurations for representative points in the ($\Gamma,\delta$) phase diagram. Snapshots are taken after the system has reached the steady orbiting state, $t\gtrsim 10^4 \, T_0$, for $\delta=0.4, \, 1.4$ and $\Gamma=-0.5231, \, 0.5231$. Particles are coloured by their orientation. All simulations are performed at $\phi=0.005$ with $\sigma_0/(\mu L_0^2U_0)=-14.95$. The particles remain well separated in all cases, suggesting suppressed large-scale density fluctuations and motivating the hyperuniformity analysis. Nematic ordering is observed in panel (a), for $\delta=0.4$ and $\Gamma=-0.5231$, and in panel (d), for $\delta=1.4$ and $\Gamma=0.5231$, consistent with the predicted ordering regimes.}}
\end{figure*}

The resulting phase map is shown in Fig.~\ref{fig:phase_diagram}(c). The global nematic order parameter is defined as 
\begin{equation}
Q = \frac{1}{N} \left| \sum_{j=1}^{N} \exp\left(i 2 \theta^{(j)}\right) \right|.
\end{equation}
Two regions exhibit clear nematic ordering, illustrated by the representative snapshots in Fig.~\ref{fig:snapshots}. Strikingly, these regions coincide with the parameter regimes in which the pair theory
predicts stable phase locking at
\begin{equation}
    \Delta=0,\pi .
\end{equation}
For the pusher swimmers considered here, $\sigma_0<0$, this condition is equivalent to
\begin{equation}
    \Gamma(1-\delta)<0.
\end{equation}
Thus nematic order appears for oblate pushers when $\delta<1$, and for prolate pushers when $\delta>1$, precisely as anticipated from the two-body phase equation.
This agreement gives a simple interpretation of the collective state. The nematic ordering does not arise merely because the particles are elongated and crowded, as in ordinary rod or ellipsoid suspensions. Its origin is hydrodynamic. The stresslet flow generated by one swimmer creates a rate-of-strain field that rotates nearby swimmers towards either parallel or anti-parallel orientation. The pair-level phase locking therefore survives at the collective scale, where it appears as global nematic alignment. The role of shape anisotropy is still evident: particles with larger $|\Gamma|$ respond more strongly to the rate of strain and reach the nematic state more rapidly, as shown in Fig.~\ref{fig:nematic}. But the mechanism is not steric packing. This is especially clear for prolate pushers, which develop nematic order only when the hydrodynamic non-axisymmetry is strong enough, namely when $\delta>1$.

We now examine the complementary regions of the phase diagram, where nematic alignment is absent. The pair theory predicts that in these regions the stable phase-locked states should instead be
\begin{equation}
    \Delta=\frac{\pi}{2},\frac{3\pi}{2}.
\end{equation}
To test this, we compute local phase-locking measures from neighboring swimmer pairs. For each adjacent pair $(i,j)$, we define
\begin{equation}
    \Delta_{ij}=\theta_j-\theta_i,
\end{equation}
and measure
\begin{equation}
    C_{\parallel}
    =
    \left\langle |\cos\Delta_{ij}| \right\rangle_{\mathrm{nn}},
    \qquad
    C_{\perp}
    =
    \left\langle |\sin\Delta_{ij}| \right\rangle_{\mathrm{nn}} .
\end{equation}
Here $\langle\cdot\rangle_{\mathrm{nn}}$ denotes an average over neighboring pairs. The quantity $C_{\parallel}$ is large when neighbors are aligned or anti-aligned, whereas $C_{\perp}$ is large when they are locked near $\pi/2$ or $3\pi/2$. As shown in Fig.~\ref{fig:phase_diagram}(b),
$C_{\perp}$ peaks precisely in the regions where the pair theory predicts the quarter-shifted states to be stable. Comparing Fig.~\ref{fig:phase_diagram}(a) and Fig.~\ref{fig:phase_diagram}(c), we see that the regions of high local nematic (as measured by $C_{\parallel}$) and global nematic order (as measured by $Q$) are roughly co-extensive, with a slightly larger region of high $C_{\parallel}$. These results further support the interpretation that collective organization is established first at the pair level, and then carried upward by  hydrodynamic interactions.

\subsection{Field Theory}

To further understand the emergence of orientational order at the collective scale, we now move from the particle-level description to a coarse-grained, continuum description. Here, we closely follow and adapt the coarse-graining approach of Das \textit{et al.}  \cite{das2024flocking}. Detailed derivations are given in Appendix C. The main result of these derivations is an expression for the growth rate of the nematic order parameter:
\begin{equation}
\begin{split}
    a_{\mathrm{nem}}
    =
    \frac{\rho}{\pi}
    \int_0^\infty dr\,r
    \int_0^{2\pi}d\phi
    \int_0^{2\pi}d\theta\;
    g(r,\phi,\theta)
    \omega(r,\phi,\theta)
    \sin 2\theta .
\end{split}
\label{eq:nematic_growth_final}
\end{equation}
Here, $g(r,\phi,\theta)$  is the pair distribution function and $\omega(r, \phi, \theta)$ encodes the pairwise rotation of one swimmer by another as a function of their relative spatial separation and orientation (see Appendix C for definition of the coordinates). The isotropic state is unstable to nematic order when
$a_{\mathrm{nem}}>0$.  The integral in Eq.~\eqref{eq:nematic_growth_final} can be determined using the interaction term in Eq. \eqref{eq:jeffery} for $\omega(r,\phi,\theta)$ if the pair distribution $g(r,\phi,\theta)$ is known. As in the approach of Das \textit{et al.}, this function may be measured directly from simulations and used as an input to the theory \cite{das2024flocking}. This retains the pair correlations generated by the microscopic dynamics.

We also consider the mean-field limit, in which these correlations are neglected: $g(r,\phi,\theta) = 1$. In this approximation, we can obtain an analytical expression for the growth rate: 
\begin{equation}
    a_{\mathrm{nem}}^{\mathrm{MF}}
    =
    \frac{3\rho\Gamma\sigma_0(1-\delta)}
    {32\mu r_{\min}} 
    \label{eq:nematic_growth_mf_final},
\end{equation}
where $r_{\min}$ is a cutoff distance. Thus, within the mean-field approximation, the isotropic state becomes unstable
to nematic order when $a_{\mathrm{nem}}^{\mathrm{MF}}>0$. For the pusher swimmers studied here, $\sigma_0<0$, this occurs when
\begin{equation}
    \Gamma(1-\delta)<0,
\end{equation}
or, equivalently, in either of the two regimes
\begin{equation}
    \Gamma<0,\ \delta<1,
    \qquad
    \Gamma>0,\ \delta>1 .
\end{equation}
The same expression also shows that the instability strengthens with increasing
$|\Gamma|$: more strongly anisotropic particles develop nematic order on a
shorter time scale. This prediction agrees well with the simulations, as shown
in Fig.~\ref{fig:nematic}. 

\subsection{Steady-state structure}

We next examine the structure of the steady states associated with the global synchro-nematic and local synchro-antinematic regimes. This question is natural in the present system, since chirality can suppress long-wavelength density fluctuations, and may therefore give rise to exotic spatial arrangements such as hyperuniform active states \cite{huang2021circular}. The procedure used to identify the steady state is described in Appendix D.

The $(\delta,\Gamma)$ phase diagram is divided into four distinct stability sectors by the conditions in Eqs. ~\eqref{eq:phase_stability_parallel} and \eqref{eq:phase_stability_quarter}. To sample the structural behavior across this phase space, we choose one representative point from each sector:
(i) \(\Gamma=0.5231,\ \delta=0.4\), synchro-antinematic prolate pushers;
(ii) \(\Gamma=-0.5231,\ \delta=0.4\), synchro-nematic oblate pushers;
(iii) \(\Gamma=0.5231,\ \delta=1.4\), synchro-nematic prolate pushers; and
(iv) \(\Gamma=-0.5231,\ \delta=1.4\), synchro-antinematic oblate pushers.

The steady-state snapshots (Fig. \ref{fig:snapshots}) already show that the swimmers are not randomly placed. They are well separated and, locally, appear to form fairly regular neighbor cages. In two dimensions, such local packing naturally suggests sixfold order. We therefore begin by measuring the local hexatic order parameter, 
\begin{equation}
\psi_{6,j} = \frac{1}{N_j} \sum_{k=1}^{N_j} \exp(i6\theta_{jk}), 
\end{equation}
where the sum is over the Voronoi neighbors of swimmer $j$, and $\theta_{jk}$ is the angle made with the $x$ axis by the line joining swimmer $j$ to neighbor $k$. This quantity distinguishes a merely well-spaced configuration from one in which each swimmer is surrounded by an approximately hexagonal cage.
\begin{figure*}[hbtp]
    \centering
    \includegraphics[width=\textwidth]{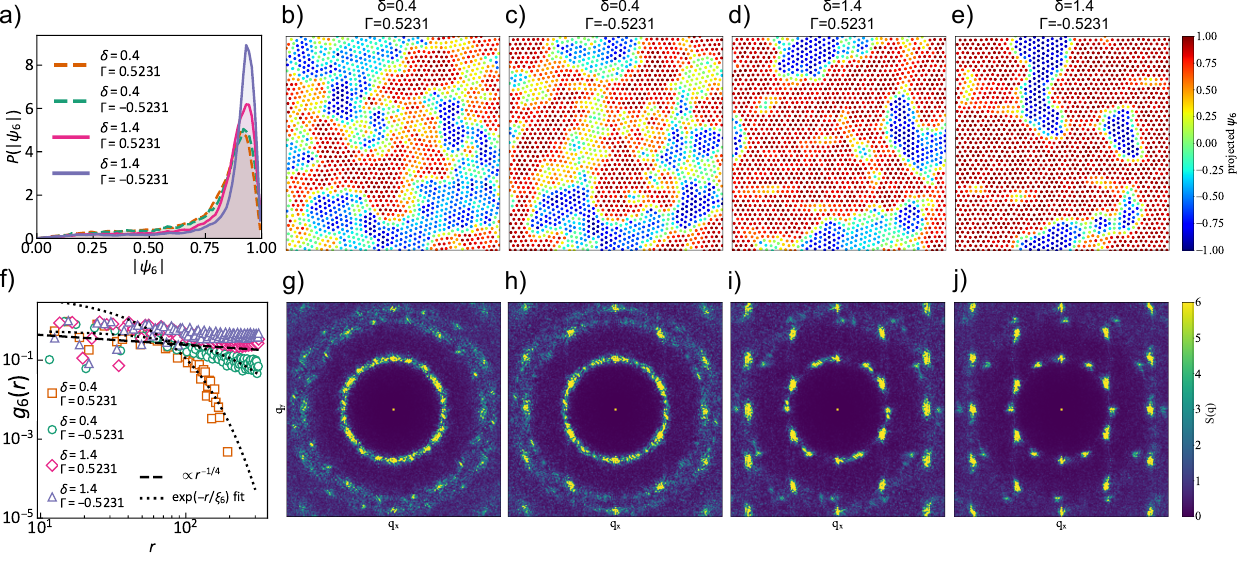}
    \caption{\label{fig:collective}\justifying{(a) Probability distribution $P(|\psi_6|)$ for the magnitude of the local hexatic order parameter $\psi_6$ for $\delta = 0.4, 1.4$ and $\Gamma = -0.5231, 0.5231$. (b-e) Projected local hexatic order parameter $\psi_{6}^{proj}$ for the same values of $\delta$ and $\Gamma$. This quantity measures how well the local hexatic order agrees with the global mean $\Psi_6$, and its spatial variation shows the orientational domain structure of the system.  (f) Decay of the bond-orientational correlation function as a function of distance (in units of $L_0$). The dotted curves are exponential fits. The dashed line shows a $r^{-1/4}$ power law for comparison with the data. (g)-(j) Two-dimensional static factors $S(q_x, q_y)$ for the same values of $\delta$ and $\Gamma$. }}
\end{figure*}
The probability distributions $P(|\psi_6|)$, shown in Fig.~\ref{fig:collective}(a), are strongly peaked close to $|\psi_6|=1$ for all four states. Thus, all four systems possess pronounced local sixfold packing. The peaks are sharper for $\delta=1.4$, indicating that the local hexagonal environments are not only present, but more uniformly developed across the sample. This, however, is only a statement about local order. The magnitude $|\psi_6|$ says little about whether neighboring cages share the same orientation.

To address this, we plot the projected local hexatic order. We first define the mean hexatic orientation of the system, 
\begin{equation} 
\Psi_6 = \frac{1}{N} \sum_{j=1}^{N} \psi_{6,j}, \qquad \hat{\Psi}_6 = \frac{\Psi_6}{|\Psi_6|}, 
\end{equation}
and then project each local value onto this direction:
\begin{equation} 
\psi^{\mathrm{proj}}_{6,j} = \mathrm{Re} \left[ \psi_{6,j}\hat{\Psi}_6^{*} \right]. \end{equation}
Positive values correspond to local hexagonal cages aligned with the mean orientation, while negative values correspond to cages rotated relative to it. The projected maps therefore reveal the domain structure hidden by the scalar distribution $P(|\psi_6|)$. The projected $\psi_6$ maps (Fig.~\ref{fig:collective}(b-e)) show a clear distinction between the two values of $\delta$. For $\delta=0.4$, the system is divided into several orientational domains, with no single hexatic orientation dominating the whole sample. For $\delta=1.4$, by contrast, the maps are largely controlled by one color, especially for $\Gamma=-0.5231$. Most local hexagonal cages then share a common orientation, with only a few smaller rotated patches. Thus, increasing $\delta$ appears to promote a more coherent spatial organization of the local sixfold order.

To quantify how far the local sixfold orientation persists, we compute the bond-orientational correlation function 
\begin{equation} 
g_6(r) = \frac{ \left\langle \psi_{6,i}^{*}\psi_{6,j} \right\rangle_{|\mathbf x_i-\mathbf x_j|=r} }{ \left\langle |\psi_6|^2 \right\rangle }. 
\end{equation}
Here the average is over particle pairs separated by a distance $r$. While $(P(|\psi_6|)$ measures the strength of local packing, $g_6(r)$ measures the range over which the orientation of that packing remains correlated. In two-dimensional melting, the distinction between liquid, hexatic, and solid phases is made precisely through the decay of orientational and positional correlations: liquids have short-ranged (exponentially decaying) orientational and positional order, hexatics have quasi-long-ranged (algebraically decaying) bond-orientational order but short-ranged positional order, and solids have quasi-long-ranged positional order together with truly long-ranged orientational order \cite{bernard2011two,digregorio2018full}. The same correlation based logic has been used to classify active two-dimensional systems, where activity shifts the phase boundaries but does not remove the relevance of the liquid--hexatic--solid distinction \cite{digregorio2018full,klamser2018thermodynamic}.

For $\delta=0.4$, $g_6(r)$ decays rapidly and is well described by an exponential form, indicating a finite bond-orientational correlation length (Fig.~\ref{fig:collective}(f)). These states are therefore liquid-like with respect to bond-orientational order. They are not, however, ordinary liquids. Their low-$q$ static structure factor (defined below) shows suppressed long-wavelength density fluctuations, so we describe them as hyperuniform liquid-like states with short-ranged hexatic correlations. The hyperuniformity analysis is given in Appendix~E. For $\delta=1.4$, $g_6(r)$ decays far more slowly and is closer to an algebraic form (Fig.~\ref{fig:collective}(f)). This is the signature expected when the system approaches hexatic-like order. 

To determine whether the orientational order at $\delta = 1.4$ is accompanied by positional crystalline order, we next examine the two-dimensional static structure factor, \begin{equation}
S(\mathbf q) = \frac{1}{N} \left\langle \left| \sum_{j=1}^{N} e^{i\mathbf q\cdot \mathbf x_j} \right|^2 \right\rangle, 
\label{eq:sq2d}
\end{equation}
where \(\mathbf q=(q_x,q_y)\), and the average is taken over steady-state configurations. For $\delta=0.4$, $S(\mathbf q)$ has a diffuse ring-like form (Fig.~\ref{fig:collective}(g,h)) with weak azimuthal modulation. This indicates a preferred interparticle spacing and local, but not long-ranged, bond-orientational order. Such a pattern is consistent with liquid-like or finite-domain hexatic order. As mentioned, $S(\mathbf q)$ vanishes as $|\mathbf{q}| \rightarrow 0$, indicative of hyperuniformity. (At the single point $\mathbf{q} = 0$, $S(0) = N$ by definition.)  For $\delta=1.4$, the first diffraction shell develops sharper Bragg-like spots, suggesting stronger positional ordering and more extended crystalline-like domains (Fig.~\ref{fig:collective}(i,j)). The first shell contains roughly twelve spots rather than the six expected for a hexagonal lattice. One possible interpretation is that two dominant hexagonal domains with different orientations contribute two sixfold sets of reciprocal-lattice peaks. This agrees with the projected $\psi_6$ maps, where the $\delta=1.4$ states show a dominant orientation together with smaller rotated domains.  Finite system size, limited domain size, or peak splitting may also contribute to the apparent multiplicity of peaks. In particular, because a hexagonal lattice is incommensurate with a square periodic simulation cell, the resulting elastic strain and defects could stabilize the multidomain structure. We therefore describe these states cautiously as hexatic- or polycrystalline-like, rather than as perfect single crystals.

Taken together, the four steady states span a structural sequence: the two $\delta=0.4$ cases remain locally hexagonal but orientationally fragmented and liquid-like, while the two $\delta=1.4$ cases develop stronger domain coherence and sharper crystalline signatures, approaching hexatic- or polycrystalline-like order.

We next ask why increasing $\delta$ appears to favour this more coherent structural organisation. A plausible origin of this trend lies in the stresslet asymmetry (Eq.~\ref{eq:non_axis_stresslet}). Increasing $\delta$ makes the entire flow field stronger. This provides a possible explanation for why the $\delta=1.4$ states show fewer orientational domains, slower decay of orientational correlations, and sharper reciprocal-space peaks.

This analysis should be read as an initial structural diagnosis, not as a complete phase classification. A definitive identification of liquid, hexatic, and solid phases would require the full machinery used in two-dimensional melting studies: finite-size scaling, careful fits of $g_6(r)$, positional correlation functions, defect statistics, Binder-type analyses, and robust low-$q$ scaling of $S(q)$ \cite{bernard2011two,digregorio2018full}. The point here is more modest, but still rather suggestive. The same chiral hydrodynamic system appears capable of producing hyperuniform liquid-like states, extended hexatic-like domains, and possibly polycrystalline solid-like order, depending on the stresslet asymmetry.

This structural tunability distinguishes the present system from earlier non-chiral squirmer monolayers. For squirmers confined to near-surface motion by gravity, hydrodynamic repulsion produces a hydrodynamic Wigner fluid with visible local hexagonal packing, but without clear long-range translational or orientational order \cite{kuhr2019collective}. Introducing an orienting field (such as bottom-heaviness) can drive the spontaneous formation of spinning dimers and trimers. A homochiral sample of these clusters can arrange into a hexagonal crystal, but a racemic mixture will stay disordered \cite{shen2019hydrodynamic}. In the present work, the interplay of chirality, particle shape, and stresslet anisotropy selects from among polycrystalline /  hexatic-like and liquid-like steady states.

\subsection{Robustness against finite hydrodynamic size and near-field hydrodynamics}

As discussed above, our kinetic theory neglects the finite hydrodynamic size of the microswimmers and near-field hydrodynamic interactions, as well as the effect of geometric confinement. Here, we seek to demonstrate that global nematic alignment can still be recovered when these effects are resolved.  We focus on oblate squirmers with $\delta = 0$ in a parallel-plate geometry, and defer consideration of prolate squirmers and antinematic ordering to a future investigation.  

\begin{figure}
    \centering
    \includegraphics[width=\linewidth]{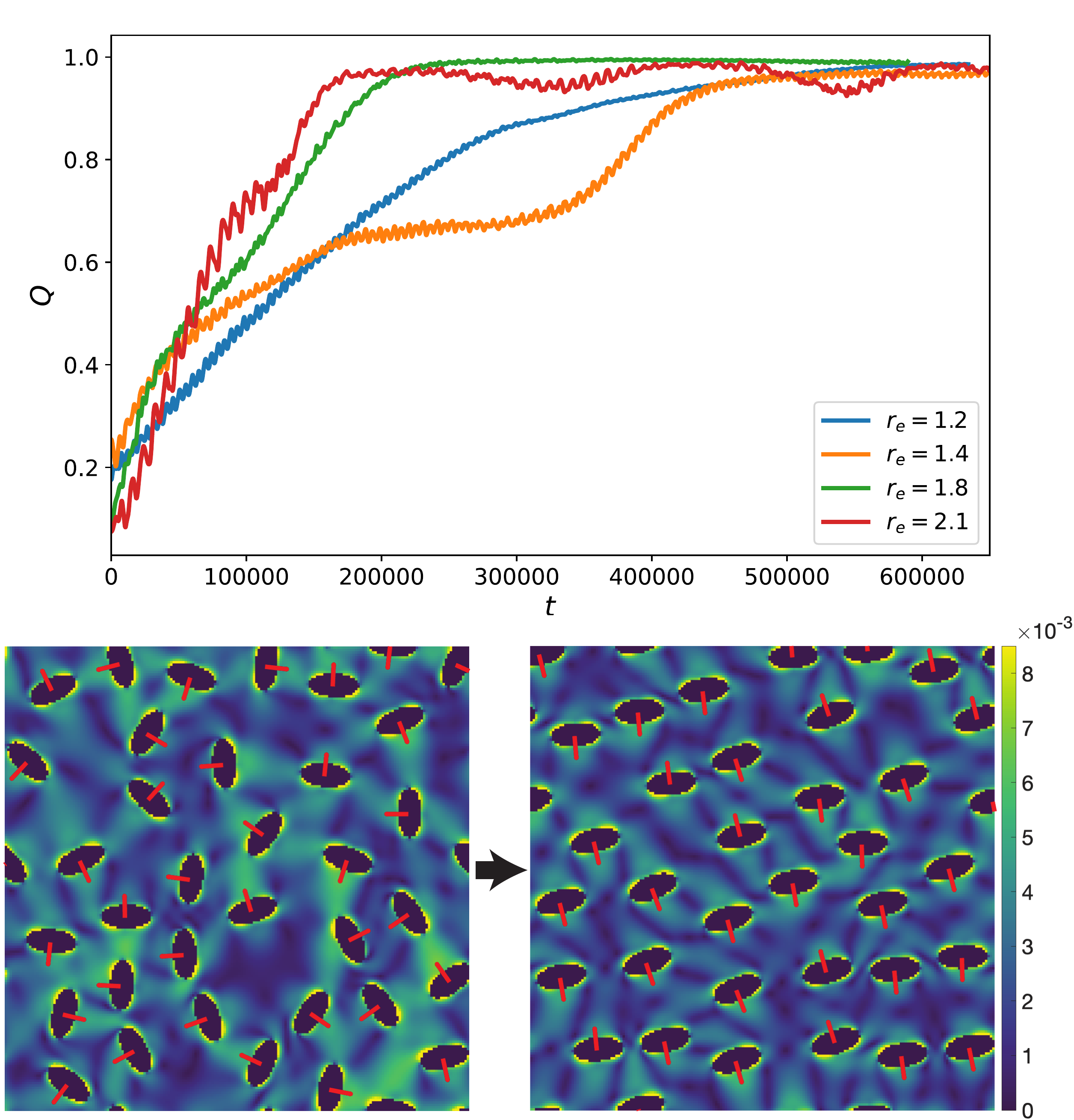}
    \caption{\justifying{(top) Global nematic order parameter $Q$ vs. LBM timestep $t$ for simulations with four different aspect ratios $r_e$. (bottom row) Initial and final states for a monolayer of oblate chiral squirmers, simulated with the Lattice Boltzmann Method. Red arrows show the particle orientation $\hat{\mathbf{d}}$. The particle aspect ratio is $r_e = 2.1$. The final state was obtained after $9.2 \times 10^{5}$ LBM steps. The background color shows the magnitude of the flow field in LBM units. 
    }}
    \label{fig:lbm_evolution}
\end{figure}

Fig. \ref{fig:lbm_evolution}, top, shows the time evolution of the nematic order parameter $Q$ for different aspect ratios $r_e > 1$. As expected, all particles achieve nematic order, and particles with larger aspect ratio order more quickly. In Fig. \ref{fig:lbm_evolution}, bottom, we show snapshots of the initial and steady-state configurations for an LBM simulation of particles with $r_e = 2.1$.

\section{Conclusions}

In summary, we have developed a theoretical framework that resolves how the interplay of three generic physical ingredients -- particle shape, long-ranged hydrodynamic interactions, and circle swimming --  can lead to synchronization at the pair level, and spatiotemporal ordering in collective phases. In particular, we find that freely swimming chiral pusher microswimmers can spontaneously phase-lock through hydrodynamic interactions. This locking may occur in parallel/antiparallel or transverse (quarter-shifted) configurations. At the collective level, the pairwise locking gives rise to global nematic alignment, in the former case, or local anti-nematic alignment, in the latter case. In these collective states, the individual particle directors rotate at the same constant rate, making them \textit{synchro-nematic} and \textit{synchro-antinematic} phases. Notably, the temporal and orientational order observed in the many-swimmer system is not imposed by an explicit alignment rule. Rather, it emerges from the hydrodynamic synchronization of swimmers whose circular motion gives them a natural phase. 

The locking mechanism is controlled by a simple parameter combination: $\Gamma(1-\delta)<0$. The sign of this expression identifies the regimes in which the pair-level hydrodynamic interaction drives neighbouring swimmers towards parallel/anti-parallel (negative) or quarter-shifted (positive) phase locking. A key point is that having a non-spherical shape ($\Gamma \neq 0)$ is essential. However, this mechanism is not restricted to one particular swimmer geometry. Oblate pusher swimmers  ($\Gamma < 0$) show parallel/anti-parallel  phase locking even in the  case of axisymmetric  actuation ($\delta=0$), corresponding to the usual squirmer-like stresslet symmetry. Prolate pushers ($\Gamma > 0$), by contrast, exhibit quarter-shifted locking when $\delta=0$, and require a non-axisymmetric stresslet for parallel/anti-parallel locking. 

The model used here is deliberately minimal. Each swimmer translates with a constant active speed, follows an intrinsic circular trajectory through a prescribed angular velocity, and interacts with other swimmers through its stresslet-generated flow field. The orientation dynamics are governed by a Jeffery-type equation, modified only by the intrinsic spin that makes the swimmer chiral. Despite this modest construction, the essential physics is captured. The strain field generated by a pusher has a nematic symmetry. As a result, a neighbouring swimmer can be hydrodynamically rotated into either a parallel/antiparallel or quarter-shifted configuration. Chirality then turns this orientational tendency into a phase-locking problem, because the particles repeatedly sample one another’s flow over their circular orbits.

This picture is supported at two levels. At the pair level, by averaging over the fast circular motion, we obtained the stable phase-locking states and their dependence on $\Gamma$ and $\delta$. At the continuum level, field theory predicts the same nematic instability regions, showing that the macroscopic nematic order is the many-body expression of the pairwise parallel/anti-parallel phase-locking mechanism. The agreement between these two descriptions is important: it shows that the collective phase diagram follows a clear hydrodynamic route from pair synchronization to bulk temporal and orientational order.

The Lattice Boltzmann simulations provide additional support to this physical picture. They allow the hydrodynamic problem to be treated with resolution of the particles' finite hydrodynamic size and near-field flows, as well as near-wall geometric confinement. As a result, we could demonstrate that the synchro-nematic ordering is robust against the presence of these effects.  Future work could extend our proof-of-concept LBM simulations to systematically study both oblate and prolate particles, as well particles with $\delta \neq 0$, for larger system sizes and in different confinement regimes.

The structural behaviour of the system adds another layer to the story. Depending on the governing parameters, the system can achieve liquid-like or polycrystalline / hexatic-like spatial order, for both the nematic and anti-nematic regimes. In all cases, the steady state exhibits strong suppression of long-wavelength density fluctuations. Both the low-q structure factor and the real-space number variance indicate class-I hyperuniformity, with scaling consistent with $S(\mathbf{q})\sim |\mathbf{q}|^2$ and $\sigma^2(R)/R^2\sim 1/R$. This suggests that orientational synchronization and spatial uniformity are governed by related but distinct aspects of the dynamics. Synchronization controls the orientational order, while the period-averaged repulsive character of coplanar circular pushers suppresses large-scale density fluctuations. In that sense, the system is ordered twice over: dynamically, through  phase locking, and structurally, through hyperuniform spatial organization.

These findings place chiral pusher suspensions in a useful position within active matter. Earlier studies have shown hydrodynamic synchronization in neutral spherical squirmers and in externally constrained or track-bound rotors. Here, we show that pusher-type microswimmers, moving freely and interacting only through their own hydrodynamic fields, can achieve the same essential outcome once particle shape and stresslet anisotropy are properly accounted for. The result is therefore directly relevant to biological and synthetic microswimmers that move on circular paths, particularly where pusher-like flows dominate.

The simplicity of the model also makes the mechanism experimentally accessible. ICEP swimmers and catalytic active particles often generate pusher-like flows quite naturally, and asymmetric designs such as L-shaped or otherwise shape-biased colloids already offer practical routes to circular motion. While experimental systems inevitably involve additional complexities that are absent from the idealized model considered here, it seems plausible that, with sufficient control over geometry and activity, signatures of the phase-locking behaviour identified in this work could be explored in experimental settings. The accompanying hyperuniformity further broadens the appeal of the system. Hyperuniform materials are known to support complete and isotropic photonic band gaps comparable to those of photonic crystals, while remaining more robust to structural disorder and fabrication imperfections. The absence of long-range periodicity also offers a route to direction-independent waveguiding and more flexible optical geometries~\cite{yu2021engineered}. 

Our work may also find application in physical reservoir computing \cite{wang2024harnessing,heuthe2026reservoir}. The input could be encoded by modulating the particles' squirming modes, and the output read from their positions and orientations. The system parameters could be tuned to position the system with respect to the phase boundaries in order to select for an optimal combination of memory (relaxation of soft collective modes) and stability. Notably, the $\sin(2 \Delta)$ dependence of the orientational interaction naturally endows the system with the nonlinearity needed for nontrivial transformation from input to output.

Taken together, the work shows that chirality, hydrodynamics, and particle shape form a remarkably effective framework for spatiotemporal collective order. Chirality gives each swimmer a phase, hydrodynamics allows one swimmer to tune the phase and orientation of another, and particle shape decides whether that tuning is stabilizing or not. From these ingredients alone, freely moving pusher microswimmers can pass from individual circular motion, to pairwise phase-locking, to global nematic or local anti-nematic order, while simultaneously arranging themselves into a crystalline or disordered hyperuniform spatial structure. This constitutes the main message of the study.
\bibliography{circle_swimmers} 

\section{Acknowledgments}
 This research was  sponsored by the Army Research Office and was accomplished under Grant Number W911NF-23-1-0190. The views and conclusions contained in this document are those of the authors and should not be interpreted as representing the official policies, either expressed or implied, of the Army Research Office or the U.S. Government. The U.S. Government is authorized to reproduce and distribute reprints for Government purposes notwithstanding any copyright notation herein. The technical support and advanced computing resources from University of Hawaii Information Technology Services – Cyberinfrastructure, funded in part by the National Science Foundation CC* awards \#2201428 and \#2232862 are gratefully acknowledged. This work used Bridges-2 at Pittsburgh Supercomputing Center through allocation PHY250064 from the Advanced Cyberinfrastructure Coordination Ecosystem: Services \& Support (ACCESS) program, which is supported by National Science Foundation grants \#2138259, \#2138286, \#2138307, \#2137603, and \#2138296. W.E.U. gratefully acknowledges the hospitality of SKCM\textsuperscript{2} during his sabbatical stay in Fall 2025, where this work was initiated. A.T. thanks Ricard Alert for helpful discussion concerning field theory.

\section{Appendix A: Details and validation for oblate squirmers in the LBM }

We consider an oblate spheroidal squirmer with its geometric centroid at the origin of a cartesian coordinate system $(\tilde{x},\tilde{y},\tilde{z})$. The minor axis of the spheroid is aligned with the $\tilde{z}$-axis. The oblate spheroidal coordinate system is given by
\begin{equation}
\lambda = \frac{1}{2 \bar{c}} \left( \sqrt{\tilde{x}^2 + \tilde{y}^2 + (\tilde{z} - i \bar{c})^2 } + \sqrt{\tilde{x}^2 + \tilde{y}^2 + (\tilde{z} + i \bar{c})^2 } \right)
\end{equation}
\begin{equation}
\zeta = -\frac{1}{2 i \bar{c}} \left(   \sqrt{\tilde{x}^2 + \tilde{y}^2 + (\tilde{z} - i \bar{c})^2 } - \sqrt{\tilde{x}^2 + \tilde{y}^2 + (\tilde{z} + i \bar{c})^2 } \right)
\end{equation}
\begin{equation}
\phi = \arctan \left(\frac{\tilde{y}}{\tilde{x}}\right)    
\end{equation}
with $\bar{c} = \sqrt{b_x^2 - b_z^2}$. In our implementation of LBM, a point at the solid/fluid interface in LBM coordinates $(x, y, z)$ is transformed into the particle-centered coordinate system $(\tilde{x},\tilde{y},\tilde{z})$, and the spheroidal coordinates $(\lambda, \zeta, \phi)$ are determined  from  $(\tilde{x},\tilde{y},\tilde{z})$. The slip velocity is then calculated from Eq. \eqref{eq:vs_LBM}.

We briefly describe the Lattice Boltzmann Method and the solid/fluid coupling scheme. We consider a cubic lattice of LBM nodes. At each fluid node $\mathbf{x}_j$, there are nineteen scalar populations $f_i(\mathbf{x}_j,t)$ with $i \in \{0, .., 18\}$. Each population is associated with a lattice vector $\mathbf{e}_{i}$. In each timestep, these populations evolve according to the lattice Boltzmann equation: $f_i(\mathbf{x}_j + \mathbf{e}_i \delta t,t + \delta t) = f_i(\mathbf{x}_j, t) +  [f_{i,eq}(\mathbf{x}_j, t) - f(\mathbf{x}_j, t)]/\tau$, where $\tau$ is the relaxation time. We choose a relaxation time $\tau = 1$, which sets the fluid kinematic viscosity as $\nu = 1/6$ in LBM units. At a fluid node, the density and velocity can be calculated from the populations $f_i$. Expressions for the density $\rho(\mathbf{x}_j, t)$, velocity $\mathbf{u}(\mathbf{x}_j, t)$, lattice vectors $\mathbf{e}_{i}$, and equilibrium populations $f_{i,eq}(\mathbf{x}_j, t)$ are cataloged in Ref. \citenum{aidun2010lattice}.

 In order to couple the solid squirming particles and the LBM fluid, we distinguish solid and fluid domains. LBM nodes that are inside the boundaries of a squirming particle are considered to be ``solid'' and have $f_i = 0$ for all $i$. When a population $f_i$ is streamed from a fluid node to a solid node, it is returned to the originating fluid node with reversed direction in a ``bounce-back'' scheme. The population is adjusted to account for the velocity of the solid at the fluid/solid interface. This bounce-back scheme leads to an exchange of momentum between the fluid and the particle. The force and torque on the particle are calculated from the difference between the outgoing and incoming fluid momenta. To better resolve the curvature of the solid particles, we use the  bounce-back linear interpolation scheme of Bouzidi \textit{et al.} \cite{bouzidi2001momentum}. The positions and orientations of the particles evolve according to Newtonian rigid body dynamics. Unit quaternions are used to track the orientation of each particle, and a rotation matrix connecting the world and body frame is calculated from the orientational quaternion \cite{theers2016modeling}. To prevent close contact of particles, we include short-ranged soft repulsive forces between particles. During the time evolution of the system, nodes change from solid to fluid and \textit{vice versa} as the particles move over the LBM grid. This requires removal and refill of fluid populations. In our simulations, we did not find the choice of refill/removal scheme to have a significant effect on the dynamics.

In the simulations in the main text, we set $B_1 = 0.003$, $B_2 = -0.012$, and $B^* = 0.005$. All other mode amplitudes are set to zero. The particle dimensions are $b_z = 5.2$ and $b_x = b_y = r_e \, b_z$ in lattice units, with an aspect ratio $r_e = b_x/b_z$. We confine $N_s = 36$ particles in the x-direction between parallel no-slip solid walls, represented by stationary solid nodes. In the y- and z-directions, we impose boundary conditions. The dimensions of the simulation box, in terms of number of LBM nodes in each direction, are $(l_x, l_y, l_z) = (192, 192, 192)$, with solid nodes at $x = 0$ and $x = 191$. For each particle, the minor axis is oriented parallel to the yz plane. The x-position of each particle is set to $x = 11$, which ensures that there is lubricating LBM fluid between the particle and the bottom solid nodes. The initial y- and z-positions, as well as the in-plane orientations of the minor axes, are chosen randomly, with overlapping particle placements rejected. For realization of the $B^*$ mode, we choose the plane containing the particle minor axis and the x-direction for breaking of the mirror symmetry of the $B_1$ slip velocity. This leads to circling motion of the particles around the x-direction. We can also calculate the Reynolds number in our simulations. Taking $r_e = 2$ as a typical aspect ratio, the single particle velocity in LBM (Fig. \ref{fig:LBM_speed_plot}) can be used to estimate a particle Reynolds number $\textrm{Re}_p = U_s b_x/\nu$ as $\textrm{Re}_p \lesssim 0.1$. 

\subsection{Validation}
In order to validate our implementation of an oblate squirmer in the LBM, we consider the speed of an oblate squirmer as a function of aspect ratio $r_e$. The self-propulsion speed can be calculated analytically \cite{popescu2010phoretic,poehnl2020axisymmetric} as 
\begin{equation}
\label{eq:speed_squirmer}
U_s = -\frac{\lambda_0}{2} \int_{-1}^{1} \sqrt{\frac{1 - \zeta^2}{\lambda_0^2 + \zeta^2}} \, v_s(\zeta) \, d\zeta,
\end{equation}
where $\lambda_0 = \sqrt{b_x^2 - \bar{c}^2}/\bar{c}$. Assuming $B_n = 0$ for $n \geq 3$, it follows from Eq. \eqref{eq:speed_squirmer} that $U_s$ is proportional to $B_1$.

\begin{figure}[ht]
\centering
    \includegraphics[width=0.95\linewidth]{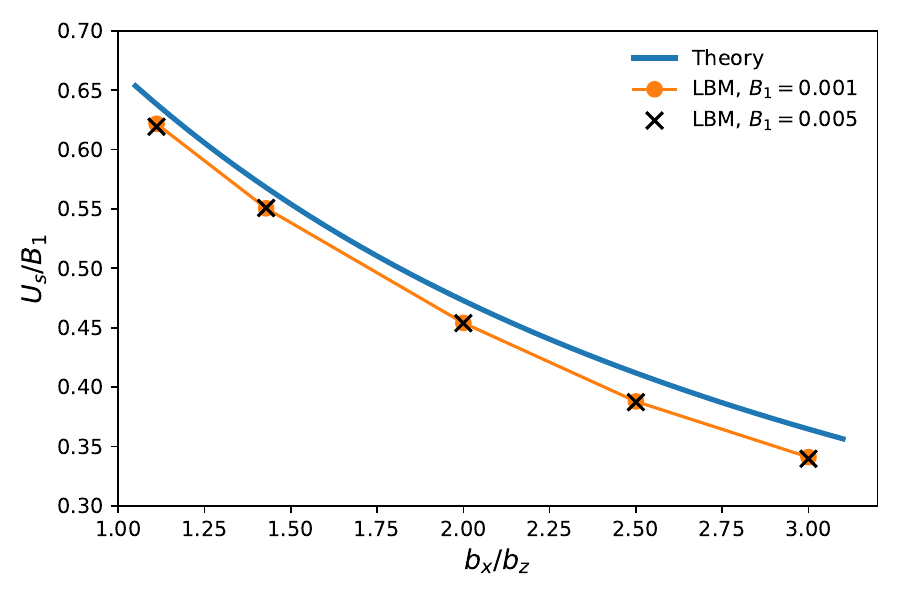}
    \caption{\label{fig:LBM_speed_plot} Normalized self-propulsion speed $U_s/B_1$ as a function of aspect ratio for an oblate spheroidal squirmer, obtained analytically (blue curve) and in the LBM model (orange and black points). The orange curve is to guide the eye. }
    
\end{figure}

Fig. \ref{fig:LBM_speed_plot} shows a comparison of $U_s/B_1$ obtained in the LBM model and the speed predicted by Eq. \eqref{eq:speed_squirmer}. The LBM squirmer is placed at position $x = l_x/2$ in a LBM cell of size $(l_x, l_y, l_z) = (128, 128, 128)$ bounded in the x-direction by parallel solid walls and with periodic boundary conditions in the y- and z-directions. The semi-minor axis length is fixed as $b_z = 5.2$ in lattice units. We set $B^* = 0$ and $B_n = 0$ for $n \geq 2$. (We have confirmed that these modes make negligible contribution to $U_s$, as expected.) The squirmer is placed with a random orientation in the yz plane, and the steady-state speed $U_s$ is calculated as the mean speed obtained after a transient period of at least 2500 LBM timesteps. The LBM data show good agreement with the prediction from Eq. \eqref{eq:speed_squirmer}, with increasing deviation for larger aspect ratios $r_e$. Slight quantitative deviation is expected, due to periodic boundary conditions,  confinement,  finite inertia, and lattice discretization in the LBM implementation. The collapse of the LBM data for two values of $B_1$ demonstrates that a linear relationship between $U_s $ and $B_1$ is approximately recovered for $B_1 \sim 10^{-3}$, despite the finite inertia inherent in the LBM. 

\begin{figure}
    \centering
    \includegraphics[width=0.95\linewidth]{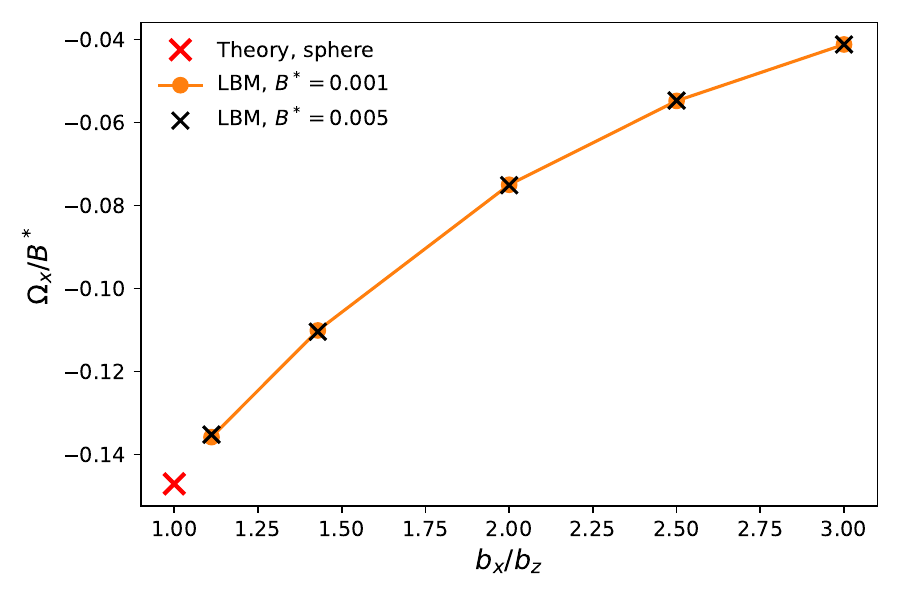}
    \caption{Normalized angular velocity $\Omega_x/B^*$ as a function of aspect ratio for an oblate spheroidal squirmer obtained with the LBM model (orange and black points). The orange curve is to guide the eye. The red point shows the analytical value for a sphere.}
    \label{fig:lbm_ox_plot}
\end{figure}

We also consider the relationship between the angular velocity and the rotational squirming mode amplitude $B^*$ (Fig. \ref{fig:lbm_ox_plot}). A linear relationship between $\Omega_x$ and $B^*$ is recovered, as expected. An analytical prediction is straightforwardly obtained for a spherical particle. Briefly, the slip velocity can be expressed in spherical coordinates as $\mathbf{v}_s = -  B^* \sgn(\sin \varphi) \sin \theta \, \hat{\theta}$. Application of the Lorentz reciprocal theorem gives \cite{stone1996propulsion} 
\begin{equation}
    \bm{\Omega} = -\frac{3}{8 \pi R^3} \int \hat{r} \times \mathbf{v}_s \, dS,
\end{equation}
yielding $\Omega_x = -\frac{3 B^*}{4 R}$ for a sphere. Notably, the angular velocity is size dependent. 

\section{Appendix B: Period-averaged vorticity and rate-of-strain}
We now analyze the velocity gradient generated by swimmer 1 in the rotating body frame of swimmer 2. The aim is to understand what part of the hydrodynamic interaction survives after averaging over one full rotation period. 

The orientation of swimmer 1 is written as
\begin{equation}
\theta^{(1)} = \beta,
\end{equation}
where $\beta$ is the fast rotational phase. 
The orientation of swimmer 2 differs from swimmer 1 by the slow phase difference $\Delta$, so that
\begin{equation}
\theta^{(2)} = \beta + \Delta.
\end{equation}
The body axes of the two swimmers are therefore
\begin{equation}
\hat{\mathbf d}^{(1)} = (\cos\beta,\sin\beta,0),
\qquad
\hat{\mathbf c}^{(1)} = (-\sin\beta,\cos\beta,0),
\end{equation}
and
\begin{equation}
\begin{aligned}
&\hat{\mathbf d}^{(2)} =
(\cos(\beta+\Delta),\sin(\beta+\Delta),0),
\\
&\hat{\mathbf c}^{(2)} =
(-\sin(\beta+\Delta),\cos(\beta+\Delta),0).
\end{aligned}
\end{equation}
The stresslet generated by swimmer 1 is taken to be
\begin{equation}
\mathbf S^{(1)}
=
\sigma_0
\left(
\hat{\mathbf d}^{(1)}\hat{\mathbf d}^{(1)}
-\frac{\bm{\mathcal{I}}}{3}
\right)
+
\sigma_0\delta
\left(
\hat{\mathbf c}^{(1)}\hat{\mathbf c}^{(1)}
-
\hat{\mathbf e}^{(1)}\hat{\mathbf e}^{(1)}
\right),
\end{equation}
with
\begin{equation}
\hat{\mathbf e}^{(1)}=(0,0,1).
\end{equation}
For the in-plane calculation, the relevant components are
\begin{equation}
S_{xx}^{(1)}
=
\sigma_0
\left(
-\frac{1}{3}
+
\cos^2\beta
+
\delta\sin^2\beta
\right),
\end{equation}
and
\begin{equation}
S_{xy}^{(1)}
=
\sigma_0(1-\delta)\sin\beta\cos\beta.
\end{equation}
Without loss of generality, we place swimmer 1 at the origin and swimmer 2 at a distance $r$ along the $x$-axis,
\begin{equation}
\mathbf x^{(1)}=(0,0,0),
\qquad
\mathbf x^{(2)}=(r,0,0).
\end{equation}
The velocity field generated by the stresslet has the form
\begin{equation}
\mathbf u(\mathbf r)
=
-\frac{3}{8\pi\mu}
\frac{\mathbf r\left(\mathbf r\cdot\mathbf S_1\cdot\mathbf r\right)}
{|\mathbf r|^5}.
\end{equation}
The velocity gradient is decomposed into its symmetric and antisymmetric parts,
\begin{equation}
\mathbf E
=
\frac{1}{2}
\left[
\nabla\mathbf u
+
(\nabla\mathbf u)^T
\right],
\end{equation}
and
\begin{equation}
\mathbf W
=
\frac{1}{2}
\left[
\nabla\mathbf u
-
(\nabla\mathbf u)^T
\right].
\end{equation}
After differentiating the stresslet velocity and evaluating at $\mathbf x^{(2)}=(r,0,0)$, the in-plane rate-of-strain tensor in the lab frame can be written compactly as
\begin{equation}
\mathbf E^{\mathrm{lab}}_{1\to2}
=
\frac{3}{8\pi\mu r^3}
\begin{pmatrix}
2S_{xx}^{(1)} & -S_{xy}^{(1)}
\\
-S_{xy}^{(1)} & -S_{xx}^{(1)}
\end{pmatrix}.
\end{equation}
The corresponding antisymmetric part is
\begin{equation}
\mathbf W^{\mathrm{lab}}_{1\to2}
=
\frac{3}{8\pi\mu r^3}
\begin{pmatrix}
0 & -S_{xy}^{(1)}
\\
S_{xy}^{(1)} & 0
\end{pmatrix}.
\end{equation}
The body-frame tensors felt by swimmer 2 are obtained by projecting onto the rotating basis
$\{\hat{\mathbf d}^{(2)},\hat{\mathbf c}^{(2)}\}$. 
\begin{equation}
\mathbf E^{\mathrm{body}}_{1\to2}
=
\mathbf Q_2^T
\mathbf E^{\mathrm{lab}}_{1\to2}
\mathbf Q_2,
\qquad
\mathbf W^{\mathrm{body}}_{1\to2}
=
\mathbf Q_2^T
\mathbf W^{\mathrm{lab}}_{1\to2}
\mathbf Q_2,
\end{equation}
where
\begin{equation}
\mathbf Q_2 =
\begin{pmatrix}
\hat{\mathbf d}^{(2)} & \hat{\mathbf c}^{(2)}
\end{pmatrix}.
\end{equation}
The period average is then taken over the fast phase $\beta$ while keeping the slow phase difference $\Delta$ fixed:
\begin{equation}
\overline{\mathbf E}_{\mathrm{body}}
=
\frac{1}{2\pi}
\int_0^{2\pi}
\mathbf E^{\mathrm{body}}_{1\to2}(\beta,\Delta)
\,d\beta,
\end{equation}
and
\begin{equation}
\overline{\mathbf W}_{\mathrm{body}}
=
\frac{1}{2\pi}
\int_0^{2\pi}
\mathbf W^{\mathrm{body}}_{1\to2}(\beta,\Delta)
\,d\beta.
\end{equation}
For the antisymmetric part, the only dependence on the fast phase is through
\begin{equation}
S_{xy}^{(1)}
=
\sigma_0(1-\delta)\sin\beta\cos\beta
=
\frac{\sigma_0(1-\delta)}{2}\sin(2\beta).
\end{equation}
Therefore, the body-frame vorticity tensor is proportional to $\sin(2\beta)$, and its average over a full period vanishes:
\begin{equation}
\overline{\mathbf W}_{\mathrm{body}}
=
\mathbf 0.
\end{equation}
The period-averaged rate-of-strain tensor in the body frame of swimmer 2 is
\begin{equation}
\begin{aligned}
&\overline{\mathbf E}_{\mathrm{body}}
=
\frac{\sigma_0}{64\pi\mu r^3}\times\\
&\begin{pmatrix}
2+6\delta+3(1-\delta)\cos(2\Delta)
&
-3(1-\delta)\sin(2\Delta)
\\
-3(1-\delta)\sin(2\Delta)
&
2+6\delta-3(1-\delta)\cos(2\Delta)
\end{pmatrix}.
\end{aligned}
\end{equation}

This can be decomposed into an isotropic part and a deviatoric part.
\begin{equation}
\begin{aligned}
\overline{\mathbf E}_{\mathrm{body}}
=&
\frac{\sigma_0(1+3\delta)}
{32\pi\mu r^3}
\bm{\mathcal{I}}
+\\
&\frac{3\sigma_0(1-\delta)}
{64\pi\mu r^3}
\begin{pmatrix}
\cos(2\Delta) & -\sin(2\Delta)
\\
-\sin(2\Delta) & -\cos(2\Delta)
\end{pmatrix}.
\end{aligned}
\end{equation}
The first term is isotropic. It stretches or compresses all in-plane directions equally. It therefore does not determine a preferred extension or compression axis, and it also does not contribute to the Jeffery rotation term because
\begin{equation}
\hat{\mathbf c}_2\cdot \mathbf I \cdot \hat{\mathbf d}_2
=
\hat{\mathbf c}_2\cdot \hat{\mathbf d}_2
=
0.
\end{equation}
The relevant part for the orientation dynamics is therefore the deviatoric tensor
\begin{equation}
\mathbf M
=
\overline{\mathbf E}_{\mathrm{body}}
-
\frac{\mathrm{Tr}(\overline{\mathbf E}_{\mathrm{body}})}{2}\mathbf I.
\end{equation}
Using the expression above,
\begin{equation}
\mathbf M
=
\frac{3\sigma_0(1-\delta)}
{64\pi\mu r^3}
\begin{pmatrix}
\cos(2\Delta) & -\sin(2\Delta)
\\
-\sin(2\Delta) & -\cos(2\Delta)
\end{pmatrix}.
\end{equation}
To identify the effective extension and compression axes, consider an arbitrary unit vector in the body frame of swimmer 2,
\begin{equation}
\mathbf v(\alpha)
=
(\cos\alpha,\sin\alpha),
\end{equation}
where $\alpha$ is measured from the $\hat{\mathbf d}_2$-axis. 
The stretching rate along this direction is
\begin{equation}
\lambda(\alpha)
=
\mathbf v^T\mathbf M\mathbf v.
\end{equation}
Substituting the expression for $\mathbf M$ gives
\begin{equation}
\lambda(\alpha)
=
\frac{3\sigma_0(1-\delta)}
{64\pi\mu r^3}
\cos\left[2(\alpha+\Delta)\right].
\end{equation}
Define
\begin{equation}
A
=
\frac{3\sigma_0(1-\delta)}
{64\pi\mu r^3}.
\end{equation}
Then
\begin{equation}
\lambda(\alpha)
=
A\cos\left[2(\alpha+\Delta)\right].
\end{equation}
The extension direction is the direction that maximizes $\lambda(\alpha)$, while the compression direction minimizes it. 
If
\begin{equation}
A>0,
\end{equation}
then the maximum occurs when
\begin{equation}
\cos\left[2(\alpha+\Delta)\right]=1.
\end{equation}
Thus
\begin{equation}
\alpha_{\mathrm{ext}}
=
-\Delta,
\qquad
\alpha_{\mathrm{comp}}
=
\frac{\pi}{2}-\Delta.
\end{equation}
On the other hand, if
\begin{equation}
A<0,
\end{equation}
then the maximum occurs when
\begin{equation}
\cos\left[2(\alpha+\Delta)\right]=-1.
\end{equation}
Therefore,
\begin{equation}
\alpha_{\mathrm{ext}}
=
\frac{\pi}{2}-\Delta,
\qquad
\alpha_{\mathrm{comp}}
=
-\Delta.
\end{equation}
This shows that the period-averaged extension and compression axes in the body frame are controlled by the sign of
\begin{equation}
A
=
\frac{3\sigma_0(1-\delta)}
{64\pi\mu r^3}.
\end{equation}
For fixed $\sigma_0$, the sign of $A$ changes when
\begin{equation}
\delta=1.
\end{equation}
For $\delta<1$ and $\delta>1$, the sign of the deviatoric strain is opposite, and the directions of extension and compression are interchanged. 
Thus, crossing $\delta=1$ swaps the effective compression and extension axes in the rotating body frame of swimmer 2.

\section{Appendix C: Field theory}

In the coarse-graining approach of Das \textit{et al}., the central object is the $N$-particle probability distribution
$\Psi_N(\mathbf{x}_1,\theta_1,\ldots,\mathbf{x}_N,\theta_N,t)$. (We use $\theta_i = \theta^{(i)}$ for notational clarity in this Appendix.) The time  evolution of $\Psi_N$ is governed by the Smoluchowski equation,

\begin{equation}
    \partial_t \Psi_N
    =
    -\sum_{i=1}^{N}
    \left(
        \nabla_i\cdot \mathbf{J}_{t,i}
        +
        \partial_{\theta_i}J_{r,i}
    \right),
    \label{eq:N_particle_smol}
\end{equation}
where the currents are
\begin{equation}
    \mathbf{J}_{t,i}=\mathbf{U}^{(i)}\Psi_N,
    \qquad
    J_{r,i}=\dot{\theta}_{i}\Psi_N .
    \label{eq:N_particle_currents}
\end{equation}
Reduction through the BBGKY hierarchy gives the one-particle equation
\begin{equation}
\begin{split}
    \partial_t \Psi_1
    =
    &-\nabla_1\cdot
    \left(
        U_s\hat{\mathbf{d}}^{(1)}\Psi_1
    \right)
    -\nabla_1\cdot
    \mathbf{J}_{t,\mathrm{int}}
    \\
    &-\partial_{\theta_1}
    \left(
        \Omega_{\mathrm{s}}\Psi_1
        +
        J_{r,\mathrm{int}}
    \right).
\end{split}
\label{eq:one_particle_smol}
\end{equation}
Here, $\Psi_1\equiv\Psi_1(\mathbf{x}_1,\theta_1,t)$. The interaction currents contain the hydrodynamic
translation and rotation induced by the other swimmers:
\begin{align}
    \mathbf{J}_{t,\mathrm{int}}
    &=
    \int d0\;
    \mathbf{u}_{0\to 1}(\mathbf{r};\theta_0)
    \Psi_2(1,0),
    \label{eq:Jt_int}
    \\[3pt]
    J_{r,\mathrm{int}}
    &=
    \int d0\;
    \omega_{0\to 1}(\mathbf{r};\theta_1,\theta_0)
    \Psi_2(1,0),
    \label{eq:Jr_int}
\end{align}
where
\begin{equation}
    \mathbf{r}=\mathbf{x}_0-\mathbf{x}_1,
    \qquad
    \int d0
    \equiv
    \int d^{(2)}\mathbf{x}_0
    \int_0^{2\pi} d\theta_0 .
\end{equation}
The hydrodynamic angular velocity
$\omega_{0\to 1}$ follows from the Jeffery response of the tagged swimmer (swimmer 1) to
the strain and vorticity generated by a neighbour (swimmer 0):
\begin{equation}
\begin{split}
    \omega_{0\to 1}
    =
    \mathbf{c}^{(1)}\cdot
    \left[
        \lambda\mathbf{E}_{0\to 1}
        +
        \mathbf{W}_{0\to 1}
    \right]
    \cdot
    \mathbf{d}^{(1)} ,
\end{split}
\label{eq:omega_kernel}
\end{equation}
with $\mathbf{E}_{0\to 1}$ and $\mathbf{W}_{0\to 1}$ evaluated at
$(\mathbf{r};\theta_0)$. 
To close the hierarchy, we write the two-particle distribution as
\begin{equation}
    \Psi_2(1,0)
    =
    \Psi_1(0)\,
    g(0\mid 1)\,
    \Psi_1(1),
    \label{eq:pair_closure}
\end{equation}
where $(1)\equiv(\mathbf{x}_1,\theta_1,t)$ denotes the tagged swimmer and
$(0)\equiv(\mathbf{x}_0,\theta_0,t)$ denotes the neighbouring swimmer. The
function $g(0\mid 1)$ is the dimensionless conditional pair distribution around
the tagged swimmer; it measures the probability of finding swimmer $0$ relative
to an uncorrelated distribution.

We now specialize to a homogeneous steady reference state, such as the isotropic
state. Homogeneity implies that pair correlations depend only on the relative
separation
\begin{equation}
    \mathbf{r}=\mathbf{x}_0-\mathbf{x}_1,
    \qquad
    r=|\mathbf{r}|,
    \qquad
    \hat{\mathbf r}=\frac{\mathbf r}{r}.
\end{equation}
Isotropy implies that there is no preferred absolute direction in the laboratory
frame. Therefore, the pair distribution can depend only on relative angular
variables. We write
\begin{equation}
    g(0\mid 1)
    \rightarrow
    g(r,\phi,\theta),
    \qquad
    \theta=\theta_0-\theta_1,
    \label{eq:g_reduced}
\end{equation}
where $\phi$ is the angle between the tagged swimmer direction and the
interparticle separation direction:
\begin{equation}
    \cos\phi=\hat{\mathbf{d}}^{(1)}\cdot\hat{\mathbf r},
    \qquad
    \sin\phi=\hat{\mathbf{c}}^{(1)}\cdot\hat{\mathbf r}.
    \label{eq:phi_def}
\end{equation}
Thus, the system is isotropic only in the global sense: there is no preferred
absolute direction. The conditional distribution around a tagged swimmer may
still depend on $\phi$ and $\theta$, because the tagged swimmer defines a local
body frame.

For the one-particle distribution, isotropy means uniformity in orientation.
With the normalization
\begin{equation}
    \rho(\mathbf{x},t)
    =
    \int_0^{2\pi}d\theta\,\Psi_1(\mathbf{x},\theta,t),
\end{equation}
the homogeneous isotropic state is
\begin{equation}
    \Psi_1(\mathbf{x},\theta,t)=\frac{\rho}{2\pi}.
\end{equation}
The angular Fourier modes of the one-particle distribution are defined by
\begin{equation}
    \Psi_{1;k}(\mathbf{x},t)
    =
    \int_0^{2\pi}d\theta\,
    e^{ik\theta}\Psi_1(\mathbf{x},\theta,t),
    \qquad
    k\in\mathbb{Z}.
    \label{eq:fourier_modes}
\end{equation}
Equivalently,
\begin{equation}
    \Psi_1(\mathbf{x},\theta,t)
    =
    \frac{1}{2\pi}
    \sum_{k\in\mathbb{Z}}
    e^{-ik\theta}\Psi_{1;k}(\mathbf{x},t).
    \label{eq:inverse_fourier}
\end{equation}
For the homogeneous isotropic state, we obtain
\begin{equation}
    \Psi_{1;0}=\rho,
    \qquad
    \Psi_{1;k}=0
    \quad
    \text{for } k\neq 0,
\end{equation}

Keeping only the lowest order in spatial gradients, the one-particle equation
reduces to
\begin{equation}
    \partial_t\Psi_1
    =
    -\Omega_{\mathrm{s}}\partial_{\theta_1} \Psi_1
    -
    \partial_{\theta_1} J_{r,\mathrm{int}}
    +
    \mathcal{O}(\nabla).
    \label{eq:lowest_grad}
\end{equation}
Using the closure for $\Psi_2$, the rotational interaction current may be
written as
\begin{equation}
    J_{r,\mathrm{int}}
    =
    \Psi_1(\mathbf{x}_1,\theta_1, t) \, \omega_{\mathrm{int}}(\mathbf{x}_1, \theta_1, t),
    \label{eq:Jr_omega_int}
\end{equation}
where $\omega_{\mathrm{int}}$ is the angular velocity induced by the surrounding
particles. At lowest order in gradients,
$\Psi_1(\mathbf{x}_0,\theta_0,t)$ in the integral defining $\omega_{int}$ is approximated as the tagged particle
position $\mathbf{x}_1$. Using
$\theta=\theta_0-\theta_1$, this gives
\begin{equation}
    \omega_{\mathrm{int}}
    =
    \frac{1}{2\pi}
    \sum_{k\in\mathbb{Z}}
    e^{-ik\theta_1}
    \Psi_{1;k}
    \kappa_{0;k}.
    \label{eq:omega_int_fourier}
\end{equation}
The coefficients
\begin{equation}
\begin{split}
    \kappa_{0;k}
    =
    \int_0^\infty dr\,r
    \int_0^{2\pi}d\phi
    \int_0^{2\pi}d\theta\;
    g(r,\phi,\theta)
    \omega(r,\phi,\theta)
    e^{-ik\theta}
\end{split}
\label{eq:kappa_def}
\end{equation}
therefore contain the hydrodynamic rotation kernel weighted by the pair
correlations.

The density is $\rho=\Psi_{1;0}$, and the complex nematic field is
\begin{equation}
    \mathcal{N}=\Psi_{1;2}.
    \label{eq:nematic_field}
\end{equation}
Projection of Eq. \eqref{eq:lowest_grad} onto the $k$th mode gives
\begin{equation}
\begin{split}
    \partial_t \Psi_{1;k}
    =
    ik\Omega_{\mathrm{s}}\Psi_{1;k}
    +
    \frac{ik}{2\pi}
    \sum_{m\in\mathbb{Z}}
    \Psi_{1;m}
    \kappa_{0;m}
    \Psi_{1;k-m}
    +
    \mathcal{O}(\nabla).
\end{split}
\label{eq:general_mode_eq}
\end{equation}
Linearising about the isotropic state, $\Psi_{1;0}=\rho$, leaves only the
$m=0$ and $m=k$ terms:
\begin{equation}
    \partial_t \Psi_{1;k}
    =
    \left[
        ik\Omega_{\mathrm{s}}
        +
        \frac{ik\rho}{2\pi}
        \left(
            \kappa_{0;0}
            +
            \kappa_{0;k}
        \right)
    \right]
    \Psi_{1;k}.
    \label{eq:linearized_k_mode}
\end{equation}
For $k=2$, using $\mathcal{N}=\Psi_{1;2}$,
\begin{equation}
    \partial_t \mathcal{N}
    =
    \left[
        2i\Omega_{\mathrm{s}}
        +
        \frac{i\rho}{\pi}
        \left(
            \kappa_{0;0}
            +
            \kappa_{0;2}
        \right)
    \right]
    \mathcal{N}.
    \label{eq:nematic_mode_eq}
\end{equation}
The intrinsic angular velocity $\Omega_{\mathrm{s}}$ rotates the phase of the
complex nematic field, whereas the growth of its amplitude is set by the
hydrodynamic interaction term. The corresponding nematic growth coefficient is
\begin{equation}
\begin{split}
    a_{\mathrm{nem}}
    =
    \frac{\rho}{\pi}
    \int_0^\infty dr\,r
    \int_0^{2\pi}d\phi
    \int_0^{2\pi}d\theta\;
    g(r,\phi,\theta)
    \omega(r,\phi,\theta)
    \sin 2\theta .
\end{split}
\label{eq:nematic_growth_final_repeat}
\end{equation}
The isotropic state is unstable to nematic order when
$a_{\mathrm{nem}}>0$. 

As discussed in the main text, the integral in Eq.~\eqref{eq:nematic_growth_final_repeat} can be determined if the pair distribution $g(r,\phi,\theta)$ is known. This function may be measured directly from simulations
and used as an input to Eq. \eqref{eq:nematic_growth_final_repeat} \cite{das2024flocking}. This retains the pair correlations
generated by the microscopic dynamics.

We also consider the mean-field limit, in which pairwise correlations are neglected: $g(r,\phi,\theta)=1$. Then Eq.~\eqref{eq:nematic_growth_final_repeat} becomes
\begin{equation}
    a_{\mathrm{nem}}^{\mathrm{MF}}
    =
    \frac{\rho}{\pi}
    \int_{r_{\min}}^\infty dr\,r
    \int_0^{2\pi}d\phi
    \int_0^{2\pi}d\theta\;
    \omega(r,\phi,\theta)\sin 2\theta ,
    \label{eq:nematic_growth_mf}
\end{equation}
where $r_{\min}$ is a short-distance cutoff, which we set as $r_{\min} = 0.5$. Evaluating the angular integrals
using the explicit form of $\omega(r,\phi,\theta)$ gives Eq. \eqref{eq:nematic_growth_mf_final} in the main text.

\section{Appendix D: Identification of steady state}

This requires some care, since the particles do not become stationary: even after relaxation, they continue to move around their circular trajectories. A raw displacement would therefore mistake persistent orbital motion for structural rearrangement. Therefore, we track whether the orbits themselves continue to drift.

For a swimmer with translational speed $U_s$ and angular velocity $\Omega_s$, the ideal orbit radius is
\begin{equation}
    R_0=\frac{U_s}{\Omega_s}.
\end{equation}
In the present simulations, $U_s=0.5$ and $\Omega_s=0.5$, giving $R_0=1$. For a particle with position
\begin{equation}
    \mathbf r_i(t)=\bigl(x_i(t),y_i(t)\bigr)
\end{equation}
and orientation
\begin{equation}
    \mathbf d^{(i)}(t)=\bigl(d_{x}^{(i)}(t),d_{y}^{(i)}(t)\bigr),
\end{equation}
we define the instantaneous orbit center as
\begin{equation}
\mathbf{C}_i(t) =     \left(
x_i(t)-R_0 d_{y}^{(i)}(t),
y_i(t)+R_0 d_{x}^{(i)}(t)
\right).
\end{equation}
\begin{figure}[ht]
\centering
    \includegraphics[width=0.75\linewidth]{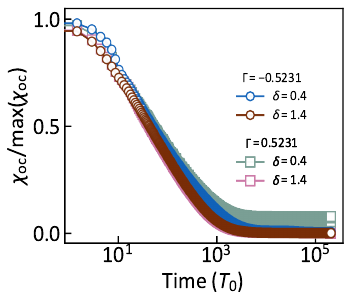}
    \caption{\label{fig:steadystate} \justifying{Orbital-centre drift $\chi_{oc}$, defined in Eq.~\eqref{eq:orbital_drift} and scaled by $R_0$, for $\delta=0.4, \, 1.4$ and $\Gamma=-0.5231,\, 0.5231$. The drift measures the displacement of each particle’s instantaneous orbit centre over a lag time $\tau$, chosen to be approximately three orbital periods. During the transient regime, $\chi_{oc}$ decreases rapidly as particles rearrange and select their orbits. For $t\gtrsim 10^4 \, T_0$, it reaches a small plateau of order $10^{-2}$, indicating that the orbit centres are nearly stationary and that the remaining dynamics is mainly persistent circular motion.}}
\end{figure}
The approach to steady state is then quantified using the orbit-center drift,
\begin{equation}
    \chi_{oc} = \frac{1}{R_0}
\sqrt{
\frac{1}{N}
\sum_{i=1}^{N}
\left|
\mathbf C_i(t+\tau)-\mathbf C_i(t)
\right|^2
}. \label{eq:orbital_drift}
\end{equation}
Here, $N$ is the number of particles and $\tau$ is chosen to be approximately three orbital periods. This choice gives the system enough time to reveal genuine orbit drift, rather than merely short-time fluctuations. The quantity $\chi_{\mathrm{oc}}$ therefore measures the root-mean-square displacement of the orbit centres over the lag time $\tau$, normalized by the orbit radius. Once the system settles, the particles continue to circulate, but the orbit centres become nearly fixed. The decay of $\chi_{\mathrm{oc}}$ to a small plateau therefore marks the point at which large-scale rearrangements have effectively ceased. From Fig.~\ref{fig:steadystate}, this occurs after approximately $10^4 \, T_0$, and all structural measurements are taken beyond this time, with additional averaging over steady-state frames and independent runs.

\section{Appendix E: Hyperuniform steady-state analysis}

We now examine the large-scale density fluctuations of the steady states. 
We first compute the full two-dimensional static structure factor $S(q_x,q_y)$ using Eq.~\ref{eq:sq2d}. As shown in Fig.~\ref{fig:collective}(g--j), the low-\(q\) density fluctuations are statistically isotropic, with no strong angular dependence around $\mathbf q=0$. This allows us to reduce $S(q_x,q_y)$ to the radially averaged structure factor $S(q)$, with $q=|\mathbf q|$, and to analyse the small-\(q\) scaling using the isotropic form described in Ref.~\cite{thambi2025clustering}.

A system is hyperuniform when its long-wavelength density fluctuations are suppressed, or equivalently when
\begin{equation}
\lim_{q\rightarrow 0} S(q) = 0.
\end{equation}
If the small-\(q\) scaling follows
\begin{equation}
S(q) \sim q^{\alpha},
\end{equation}
\begin{figure}[ht]
\centering
    \includegraphics[width=1.0\linewidth]{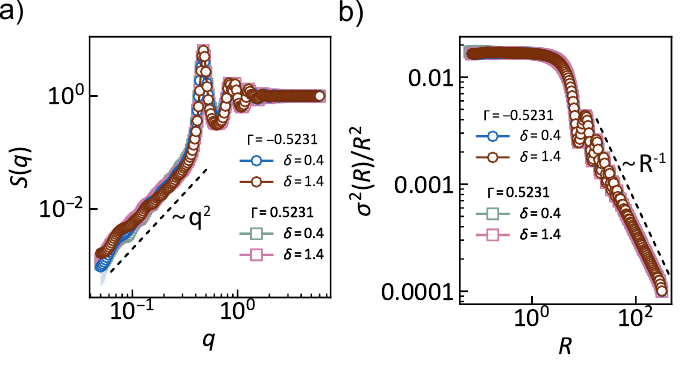}
    \caption{\label{fig:sq_var} \justifying{(a) Structure factor $S(\mathbf{q})$ for for $\delta=0.4, \, 1.4$ and $\Gamma=-0.5231,\, 0.5231$. The low-$\mathbf{q}$ behavior follows $S(\mathbf{q})\sim \mathbf{q}^2$ (black dashed line), consistent with class-I hyperuniformity. (b) Scaled number variance $\sigma^2(R)/R^2$ as a function of window radius $R$. At large $R$, the variance decays as $1/R$ (black dashed line), providing an independent real-space signature of class-I hyperuniformity.}}
\end{figure}
then \(\alpha>1\) corresponds to class-I hyperuniformity in two dimensions \cite{torquato2018hyperuniform}. For all representative steady states considered here, the low-\(q\) structure factor decays with an exponent close to \(\alpha\simeq 2\), as shown in Fig.~\ref{fig:sq_var}(a). The dark region around \(\mathbf q=0\) in the two-dimensional structure factor provides the same information visually, the longest-wavelength density modes are strongly attenuated.

The interpretation, however, is not the same in all four cases. For the $\delta=0.4$ states, $g_6(r)$ decays exponentially, identifying these systems as liquids. Their hyperuniformity is therefore a genuinely disordered form of hyperuniformity: the system lacks long-ranged or quasi-long-ranged positional order, yet still suppresses density fluctuations at large length scales. For the $\delta=1.4$ states, the situation is less surprising. These systems already show stronger orientational coherence and Bragg-like features in $S(\mathbf q)$, consistent with hexatic- or polycrystalline-like order. Since crystalline arrangements naturally suppress long-wavelength density fluctuations, their hyperuniform scaling is expected. Thus, the more notable result is not simply that all four states are hyperuniform, but that hyperuniformity persists even in the liquid-like $\delta=0.4$ regimes.

We also examine the local number variance as a real-space measure of the same effect. For a circular observation window of radius $R$, let $N(R)$ be the number of particles inside the window. The number variance is
\begin{equation}
\sigma_N^2(R)
=
\langle N^2(R)\rangle
-
\langle N(R)\rangle^2,
\end{equation}
where the average is taken over window positions, steady-state frames, and eight independent simulations. For an uncorrelated Poisson distribution in two dimensions,
\begin{equation}
\sigma_N^2(R) \sim R^2 .
\end{equation}
By contrast, a class-I hyperuniform system satisfies
\begin{equation}
\sigma_N^2(R) \sim R,
\end{equation}
or, equivalently,
\begin{equation}
\frac{\sigma_N^2(R)}{R^2}
\sim
\frac{1}{R}.
\end{equation}
The results in Fig.~\ref{fig:sq_var}(b) show this crossover clearly. At small $R$, the observation window probes only the immediate local environment, and the variance still reflects ordinary local counting fluctuations. At larger $R$, these short-scale fluctuations are averaged out, and the scaled variance approaches the expected $1/R$ decay. This real-space result is consistent with the small-$q$ scaling of $S(q)$, confirming class-I hyperuniformity.

Taken together, the reciprocal-space and real-space measurements show that the chiral steady states strongly suppress long-wavelength density fluctuations. For the more ordered $\delta=1.4$ states, this is expected, given the crystalline ordering evident in their hexatic- or polycrystalline-like structure. For the $\delta=0.4$ states, it is more distinctive. The system remains liquid-like, as determined from bond-orientational correlations, but is nevertheless hyperuniform in density fluctuations. Hence, this analysis gives the consistent physical picture that coplanar pusher swimmers interact effectively repulsively, driving the suspension towards configurations with strongly suppressed long-wavelength density fluctuations. 

\end{document}